\documentclass[conference, a4paper]{IEEEtran}
\usepackage[cmex10]{amsmath}
\usepackage{amssymb}
\usepackage{nicefrac}
\usepackage[pdftex]{graphicx}
\usepackage[tight,footnotesize]{subfigure}
\usepackage{url}
\usepackage{color}
\usepackage{ifthen}
\usepackage{balance}
\usepackage{cite}
\usepackage{epstopdf}

\IEEEoverridecommandlockouts
\begin{document}
%
%
\title{Measurement-based Online Available Bandwidth Estimation employing Reinforcement Learning}
\author{\IEEEauthorblockN{Sukhpreet Kaur Khangura and Sami Ak{\i}n}
\IEEEauthorblockA{Institute of Communications Technology\\Leibniz Universit\"{a}t Hannover}}
%
%
\maketitle
%
%
\begin{abstract}
An accurate and fast estimation of the available bandwidth in a network with varying cross-traffic is a challenging task. The accepted probing tools, based on the fluid-flow model of a bottleneck link with first-in, first-out multiplexing, estimate the available bandwidth by measuring packet dispersions. The estimation becomes more difficult if packet dispersions deviate from the assumptions of the fluid-flow model in the presence of non-fluid bursty cross-traffic, multiple bottleneck links, and inaccurate time-stamping. This motivates us to explore the use of machine learning tools for available bandwidth estimation. Hence, we consider reinforcement learning and implement the single-state multi-armed bandit technique, which follows the $\varepsilon$-greedy algorithm to find the available bandwidth. Our measurements and tests reveal that our proposed method identifies the available bandwidth with high precision. Furthermore, our method converges to the available bandwidth under a variety of notoriously difficult conditions, such as heavy traffic burstiness, different cross-traffic intensities, multiple bottleneck links, and in networks where the tight link and the bottleneck link are not same. Compared to the piece-wise linear network a model-based direct probing technique that employs a Kalman filter, our method shows more accurate estimates and faster convergence in  certain network scenarios and does not require measurement noise statistics.
\end{abstract}
%
%
%
%
\section{Introduction}\label{sec:introduction}
Real-time available bandwidth estimation in a communication network has been of interest to researchers due to its significant impact on delay sensitive Internet applications. For example, an accurate available bandwidth estimation in real-time streaming multimedia applications is fundamental  to provide certain Quality-of-Service (QoS) guarantees to end users. Furthermore, available bandwidth estimates are used to select the best route, to monitor and detect the congestion, and to balance the traffic across a network to avoid stops, lags or buffering in the streaming content. Herein, the term \emph{available bandwidth} refers to the residual capacity that remains available for data transmission after cross-traffic is served. Formally, given a link with capacity $C$ and a cross-traffic with long-term average rate $\lambda$, where $\lambda \in [0, C]$, the available bandwidth of the link is defined as $A = C - \lambda$~\cite{liu2007queueing}. Here, the end-to-end available bandwidth is determined by the \emph{tight link}, i.e., the link with the minimal available bandwidth~\cite{jain2003end}, and not by the bottleneck link, i.e., the link with the minimal capacity.

Researches have proposed several active probing techniques and corresponding theories for available bandwidth estimation~\cite{melandertopp2000new, melander2004diettopp, dovrolis2001packet, jain2002pathload, jain2003end, ribeiro2003pathchirp, strauss2003measurement, hu2003evaluation, ekelin2006real, liu2007queueing, liu2008stochastic, liebeherr2010servicecurve, luebben2014servicecurve}. These techniques use a sender that actively injects into a network  synthetic probe traffic with  known packet size, $l$, and a well-defined inter-packet gap, i.e., input gap $g_{\mathrm{in}}$.  As these probe packets traverse through the network, they get dispersed due to cross-traffic. At the receiver, this inter-packet dispersion, i.e., output gap $g_{out}$, is measured to estimate the available bandwidth in the network. These techniques have a common assumption that the cross-traffic in a network has \emph{constant-rate}. Moreover, these techniques assume a fluid-flow model and neglect the impacts of the packet granularity of the cross-traffic, i.e., the cross-traffic is assumed to be composed of infinitely  small packets. Following the constant-rate fluid-flow cross-traffic assumption, a single tight link is modeled as a lossless first-in, first-out (FIFO) multiplexer of the probe traffic and the cross-traffic. Herein, the relation between $g_{\mathrm{out}}$ and $g_{\mathrm{in}}$ is given as~\cite{liu2007queueing}
\begin{equation}\label{eqn:gap_out}
g_{\mathrm{out}} = \max \left\{g_{\mathrm{in}}, \frac{g_{\mathrm{in}} \lambda + l}{C} \right\},
\end{equation}
where $\lambda$ is the constant cross-traffic rate. Above, $g_{\mathrm{in}} \lambda$ represents the amount of the cross-traffic that enters between two probe packets having an input gap of $g_{\mathrm{in}}$, and causes them to be further apart. Reordering~\eqref{eqn:gap_out}, we particularly obtain the characteristic \emph{gap response curve} as
\begin{equation}\label{eqn:pgm}
\frac{g_{\mathrm{out}}}{g_{\mathrm{in}}} = \begin{cases}
1 & \quad \text{if } \frac{l}{g_{\mathrm{in}}} \leq C - \lambda,\\
\frac{l}{g_{\mathrm{in}}C} + \frac{\lambda}{C} & \quad \text{if } \frac{l}{g_{\mathrm{in}}} > C - \lambda. \\
\end{cases}
\end{equation}

\begin{figure}
	\centering
	\subfigure[Fluid-flow model ]{
		\includegraphics[width=0.46\columnwidth]{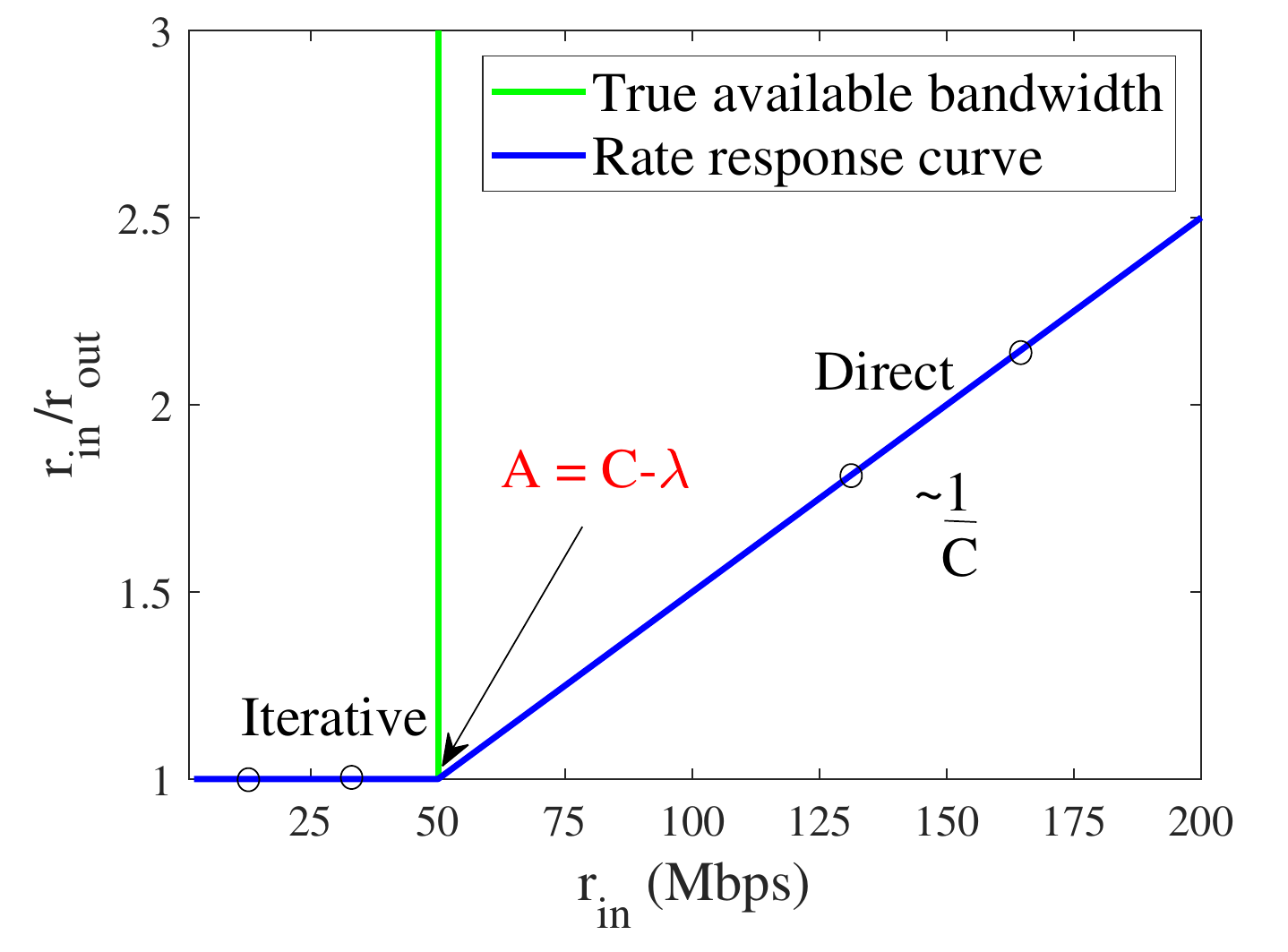}
		\label{fig:rate_res}
	}
	\subfigure[Single tight link]{
		\includegraphics[width=0.46\columnwidth]{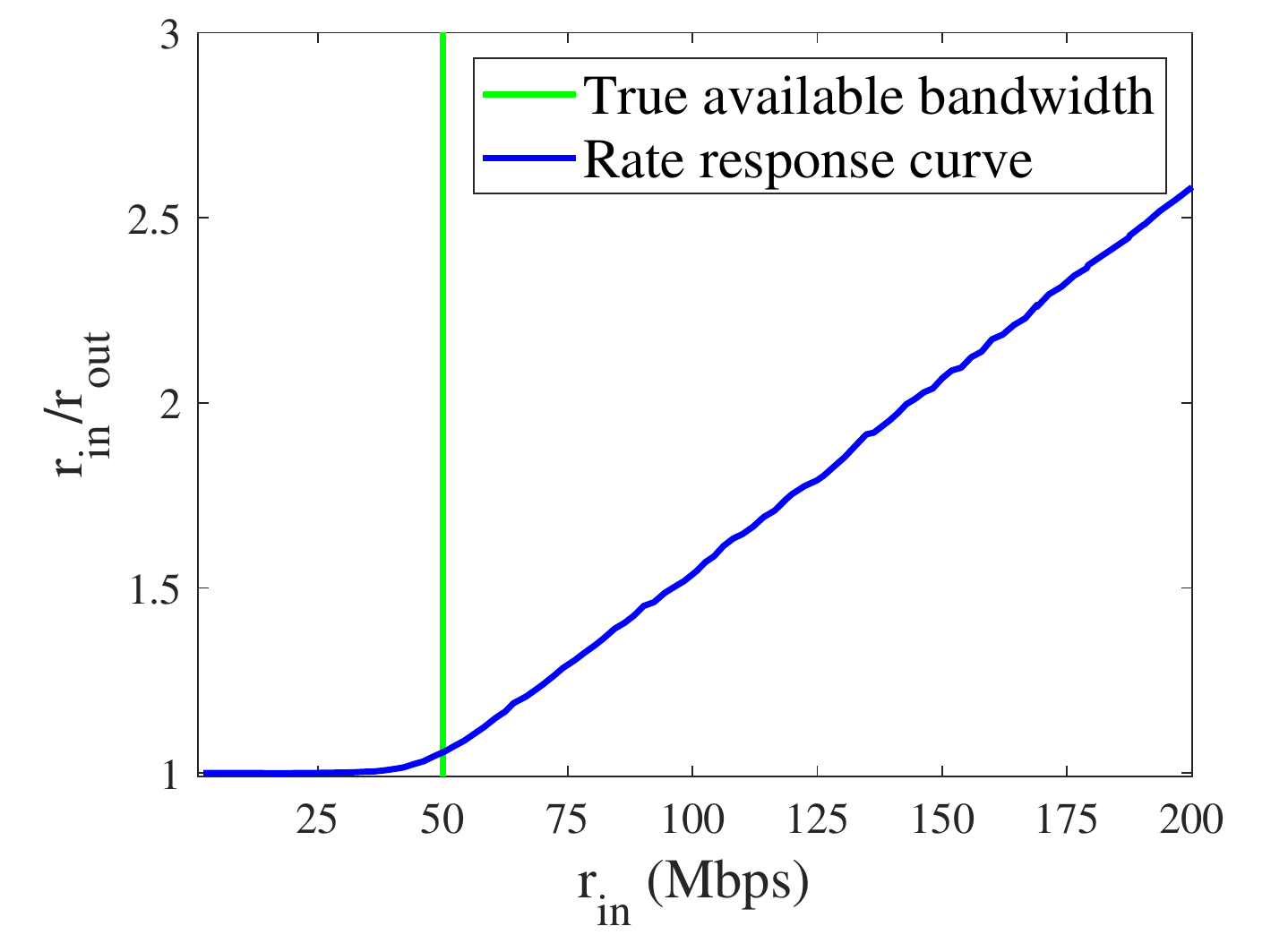}
		\label{fig:rate_res_dev}
	}
	\caption{Rate response curves of (a) the fluid-flow model assuming constant rate cross-traffic and (b) a single tight link with an exponential cross-traffic.}
\end{figure}

In practice, the  output gaps $g_{out}$ are highly distorted due to deviation from the assumptions of the model, i.e., a lossless
FIFO multiplexer with constant, fluid cross-traffic as well as
measurement inaccuracies, such as imprecise time-stamping. Therefore, the state-of-the-art bandwidth estimation methods average several output gap samples $g_{\mathrm{out}}$ in order to alleviate the observed variability of the samples of  $g_{\mathrm{out}}$. The samples can be collected by repeatedly sending \emph{packet pairs}~\cite{keshav91}, or \emph{packet trains}~\cite{paxson1999packetbunchmode, dovrolis2001packet}, which consist of $n$ consecutive packets, and hence, $n-1$ input gaps. At the receiver, the consecutive output gaps are formulated as $g_{\mathrm{out}}^{j} = t_\mathrm{out}^{j+1}-t_\mathrm{out}^{j}$ for $j = 1 \dots n\!-\!1$, where $t_\mathrm{out}^{j}$ is the time when the $j^{\text{th}}$ packet arrives at the receiver. Then, the output rate of a packet train with $n$ packets is given as
\begin{equation}\label{eq:defrout}
r_{\mathrm{out}} = \frac{(n-1)l}{t_\mathrm{out}^{n}-t_\mathrm{out}^{1}}.
\end{equation}
Since we define $g_{out}^{j} = t_{out}^{j+1} - t_{out}^{j}$, we can rewrite (3) as
\begin{equation}\label{eq:defrout_2}
r_{\mathrm{out}} = \frac{l}{\frac{1}{n-1} \sum_{j=1}^{n-1} g_{\mathrm{out}}^{j}}.
\end{equation}
Notice that the denominator in (\ref{eq:defrout_2}) converges to the mean of the output gaps with the increasing packet train size. Herein, assuming a deterministic fluid-flow model, i.e., $g_{\mathrm{out}}^{j} = g_{\mathrm{out}}$ for $\forall j$, we can see that
 \begin{equation*}
r_{\mathrm{out}} = \frac{l}{\frac{1}{n-1} \sum_{j=1}^{n-1} g_{\mathrm{out}}^{j}} =\frac{l}{\frac{(n-1) g_{\mathrm{out}}}{n-1}}= \frac{l}{g_{\mathrm{out}}}.
 \end{equation*}
Similarly, defining the input rate as $r_{\mathrm{in}} = l/g_{\mathrm{in}}$, and inserting $r_{\mathrm{in}} = l/g_{\mathrm{in}}$ and $r_{\mathrm{out}} = l/g_{\mathrm{out}}$ into ~\eqref{eqn:pgm}, we obtain the equivalent representation of the \emph{rate response curve} as
\begin{equation}\label{eqn:prm}
\frac{r_{\mathrm{in}}}{r_{\mathrm{out}}} =
\begin{cases}
1 & \quad \text{if } r_{\mathrm{in}} \leq C - \lambda,\!\\
\frac{r_{\mathrm{in}}}{C} 	+ \frac{\lambda}{C}	& \quad \text{if } r_{\mathrm{in}} > C - \lambda,\!\\
\end{cases}
\end{equation}
which mathematically describes the clear bend in the rate response curve at $r_{in} = A$ as seen in Fig.~\ref{fig:rate_res}.
%
%
\subsection{State-of-the-Art Estimation Techniques}

\begin{figure*}
	\centering
	\includegraphics[width=1.7\columnwidth]{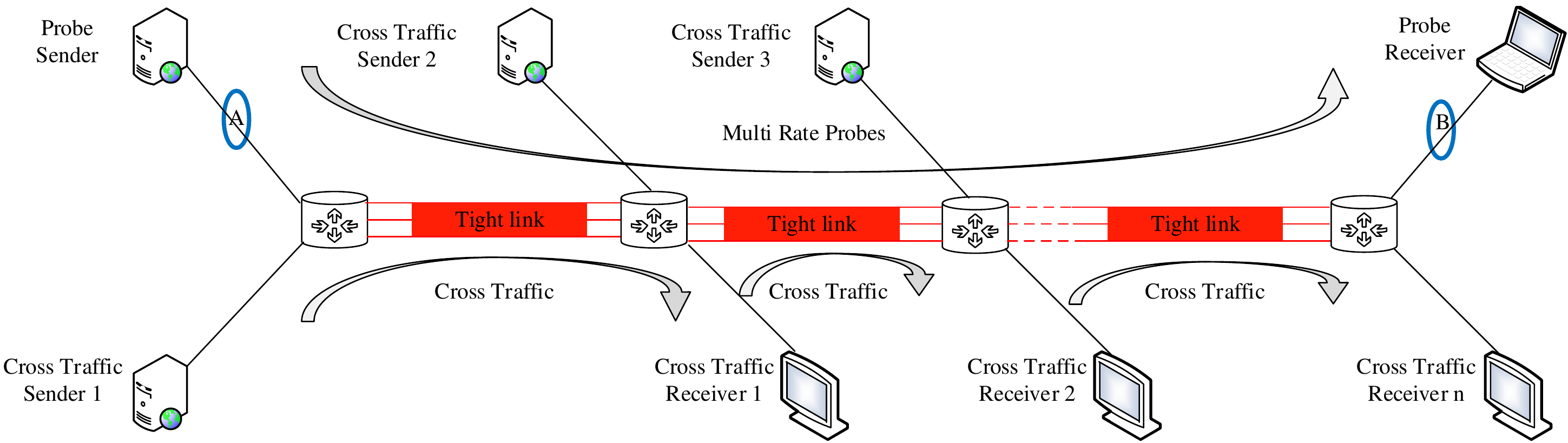}
	\caption{Dumbbell topology set up using the Emulab and MoonGen software. A varying number of tight links with single hop-persistent cross-traffic are configured. Probe traffic is path-persistent to estimate the end-to-end available bandwidth from measurements at points A and B~\cite{khangurann}.}
	\label{fig:topo}
\end{figure*}

We can classify the active available bandwidth estimation methods which are based on the constant-rate fluid-flow model as \emph{iterative probing} or \emph{direct probing} techniques. Iterative probing techniques search for the turning point in the rate response curve by sending repeated probes at increasing rates in the region defined by $r_{\mathrm{in}}/r_{\mathrm{out}}=1$. When $r_{\mathrm{in}}$ reaches $C-\lambda$, the available bandwidth saturates and increasing the probe rate to a value beyond the available bandwidth results in self-induced congestion and causes $r_{\mathrm{in}}/r_{\mathrm{out}}>1$. As a consequence, a queue builds up at the multiplexer and increases one-way delays that can be detected by the receiver. This technique is implemented, for instance, by Pathload~\cite{jain2002pathload} and Pathchirp~\cite{ribeiro2003pathchirp}. On the other hand, direct probing techniques estimate the upward segment of the rate response curve, i.e., the part where $r_{\mathrm{in}} > C-\lambda$. This segment of the rate response curve is a function of $C$ and $\lambda$. If we know $C$ a priori, we need a single probe $r_{\mathrm{in}} = C$ that yields a measurement, $r_{\mathrm{out}}$, to estimate $\lambda = C (C/r_{\mathrm{out}} - 1)$ from~\eqref{eqn:prm}. We can see the implementation of this approach in \cite{strauss2003measurement}. If we do not have the knowledge of $C$ in advance, we need at least two different probing rates $r_{\mathrm{in}} > C-\lambda$ to obtain the two unknown parameters, $C$ and $\lambda$. We can see this technique in, for instance, TOPP~\cite{melandertopp2000new}, DietTOPP~\cite{melander2004diettopp}, and BART~\cite{ekelin2006real}.

In real-time available bandwidth estimation, noisy measurement data, multiple bottleneck links, and inaccurate time-stamping degrade the estimation quality. Furthermore, the stochastic nature of cross-traffic leads to deviations from the fluid-flow model. In order to improve available bandwidth estimation and reduce the impacts of the randomness in cross-traffic, the state-of-the-art estimation methods use statistical post-processing techniques, such as a Kalman filter~\cite{ekelin2006real,bozakov09}, majority rule~\cite{jain2002pathload}, averaging repeated measurements~\cite{ribeiro2003pathchirp},~\cite{strauss2003measurement}, and linear regression~\cite{melander2004diettopp}. While packet trains and statistical post-processing techniques help to reduce the variability in available bandwidth estimation, they do not take care of the deviations from the deterministic fluid-flow model. For instance, looking at the experimental results\footnote{We obtained the results in Fig. \ref{fig:rate_res_dev} from the testbed shown in Fig.~\ref{fig:topo}. The network is set with a single tight link of capacity $C=100$~Mbps, and access links are of capacity $C=1$~Gbps. The cross-traffic is  discrete with a packet length of $l=1514$~bytes. We further set the cross-traffic to moderate burstiness with exponentially distributed packet inter-arrival times and an average rate of $\lambda=50$~Mbps. Particularly, we run the experiment 1000 times and average the results for smoothness in the presentation.} in Fig.~\ref{fig:rate_res_dev}, we can see that unlike the deterministic fluid-flow model in Fig.~\ref{fig:rate_res}, the sharp bend around $r_{\mathrm{in}} = C-\lambda$ that marks the available bandwidth is not clearly apparent. This elastic deviation from the fluid-flow model leads to biased estimates. Furthermore, it is difficult to tailor methods to specific hardware implementations that influence the measurement accuracy. Therefore, the fundamental limitations of the model-based state-of-the-art bandwidth estimation methods make researchers explore the use of machine learning techniques in bandwidth estimation.

\emph{Machine learning} techniques have taken attention initially in~\cite{eswaradass2005neural,chen2007machine} and recently in~\cite{yin2016can, sato2017experimental}. We see that the authors in~\cite{eswaradass2005neural} use network traffic data collected by passive measurements, while the authors in~\cite{chen2007machine, yin2016can, sato2017experimental} use active probes to estimate the available bandwidth in NS-2 simulations~\cite{chen2007machine}, ultra-high speed 10~Gbps networks~\cite{yin2016can}, and operational LTE networks~\cite{sato2017experimental}. Moreover, the authors in~\cite{chen2007machine, yin2016can, sato2017experimental} use packet chirps~\cite{ribeiro2003pathchirp}, i.e., the probes of several packets sent at increasing rates. They achieve the rate increase by a geometric reduction of the input gap~\cite{chen2007machine}, by concatenating several packet trains with increasing rates to a multi-rate probe~\cite{yin2016can}, and by linearly increasing the packet size~\cite{sato2017experimental}. The packet chirps, with a single probe, lead to the detection of the turning point, which is the actual available bandwidth. Nevertheless, the chirps are susceptible to random noise~\cite{liebeherr2010servicecurve}. Furthermore, the authors in~\cite{chen2007machine} study the packet bursts which are known as back-to-back packet probes and conclude that the packet bursts are not enough to estimate the available bandwidth. Also, the authors in~\cite{yin2016can} consider constant-rate packet trains in an iterative manner to attain the available bandwidth. Here, machine learning solves a classification problem to estimate whether the rate of a packet train exceeds the available bandwidth. Depending on the result, the rate of the next packet train is reduced or increased in a binary search \cite{jain2002pathload} until the probe rate approaches the available bandwidth. The authors in~\cite{yin2016can} give, however, preference to the chirp probes.

The common aspect of the machine learning implementations in available bandwidth estimation is the use of output gap measurements \cite{chen2007machine, sato2017experimental}, the Fourier transforms of input and output gaps \cite{yin2016can}, or the $k\times 1$-dimensional vectors of $r_{\mathrm{in}}/r_{\mathrm{out}}$~\cite{khangurann} as labeled input features in training and testing phases. The major objective is to invoke supervised learning to extrapolate and generalize the available bandwidth estimates for the data sets not included in the training phase. However, from a practical point of view, it is not always feasible to create such training data sets which are representative of all cases because of the bursty nature of cross-traffic and multiple bottleneck links. As a consequence of this fundamental limit in supervised learning approaches~\cite{eswaradass2005neural,chen2007machine, sato2017experimental,yin2016can,khangurann}, we motivate ourselves to use reinforcement learning in available bandwidth estimation.
%
%
\subsection{Contributions}

In this paper, we propose a method to implement reinforcement learning in available bandwidth estimation, which converges the result faster and more accurately. We consider the set of input rates as the set of actions defined in reinforcement learning theory and define a reward metric as a function of input and output rates, which reaches the maximum when the input rate is equal to the available bandwidth. Our method is different from the existing machine learning approaches used in bandwidth estimation because it does not require a training phase. We evaluate our method in a controlled network testbed, where we specifically target topologies, including a bursty cross-traffic nature and multiple bottleneck links. We consider cross-traffic scenarios with different distributions and intensities. Furthermore, we compare our method with a fluid-flow model-based direct probing technique that employs a Kalman filter and show that our method converges faster and has less variations in bandwidth estimates. Moreover, we consider more difficult scenarios by setting the tight link different from the bottleneck link. Our method reliably performs real-time available bandwidth estimation in multi-hop networks with faster convergence and less variations, where the model-based direct probing technique underestimates the available bandwidth.

The remainder of this paper is organized as follows. We describe our experimental set up in Section~\ref{sec:exp_setup} and present our reinforcement learning-based approach in Section~\ref{sec:reinforcement_based}. We introduce the reference implementation of the state-of-the-art model-based direct probing technique in Section~\ref{sec:referenceimplementation} and show test results in Section~\ref{sec:experimental_evaluation}. We provide the conclusion in Section \ref{sec:conclusion}.
%
%

\section{Experimental Setup}\label{sec:exp_setup}

We set our controlled network testbed in Leibniz Universit\"{a}t Hannover, and it comprises 80 servers. We connect each server deploying minimum four network switches with 1~Gbps and 10~Gbps link capacities. We use the Emulab software~\cite{anderson2006automatic} to manage the testbed, where we configure the servers as hosts and routers and connect them using virtual local area networks (VLANs) to implement the desired topology. We use a dumbbell topology with multiple tight links, as shown in Fig.~\ref{fig:topo}. In order to emulate the characteristics of the links, such as capacity, delay, and packet loss, we employ additional servers in Emulab. We use the MoonGen software~\cite{moongen-imc2015} for the emulation of link capacities that differ from the native physical Ethernet capacity. To achieve an accurate spacing of packets that matches the emulated capacity, we fill the gaps between packets with dummy frames by using MoonGen, which are later discarded at the output of the link. We use the ``forward rate Lua script'' for the MoonGen  to achieve the desired forwarding rate at the transmission and reception ports of MoonGen.

We create cross-traffic models having different distributions by employing distributed internet traffic generator (D-ITG)~\cite{avallone2004d}. Each cross-traffic is \emph{single-hop-persistent}, i.e., at each link, fresh cross-traffic is multiplexed. The probe traffic that we deploy to estimate the end-to-end available bandwidth is \emph{path-persistent}. Particularly, it travels the entire network path from the probe sender to the probe receiver. We use real-time user datagram protocol (UDP) data emitter and collector known as RUDE and CRUDE~\cite{laine2000real} respectively, in order to generate UDP probe streams. A probe stream consists of a series of $k$ packet trains, each having $n$ packets. These $k$ different packet trains respectively correspond to $k$ different probe rates that successively increase with an increment rate, $\delta_{r}$. We set the packet length to $l=1514$ bytes including the Ethernet header, both in the probe traffic and the cross-traffic. We use ``libpcap'' to capture the packets at the probe sender and receiver, and the packet timestamps  are generated at points A and B, respectively, as shown in Fig.~\ref{fig:topo}. We also use a specific \emph{Endace} data acquisition and generation (DAG) measurement card to obtain the accurate reference timestamps. We use the timestamps to compute $r_{\mathrm{in}}$ and $r_{\mathrm{out}}$ for each packet train.
%
%
\section{Reinforcement Learning-based Method}\label{sec:reinforcement_based}
\begin{figure}
	\centering
	\includegraphics[width=0.7\columnwidth]{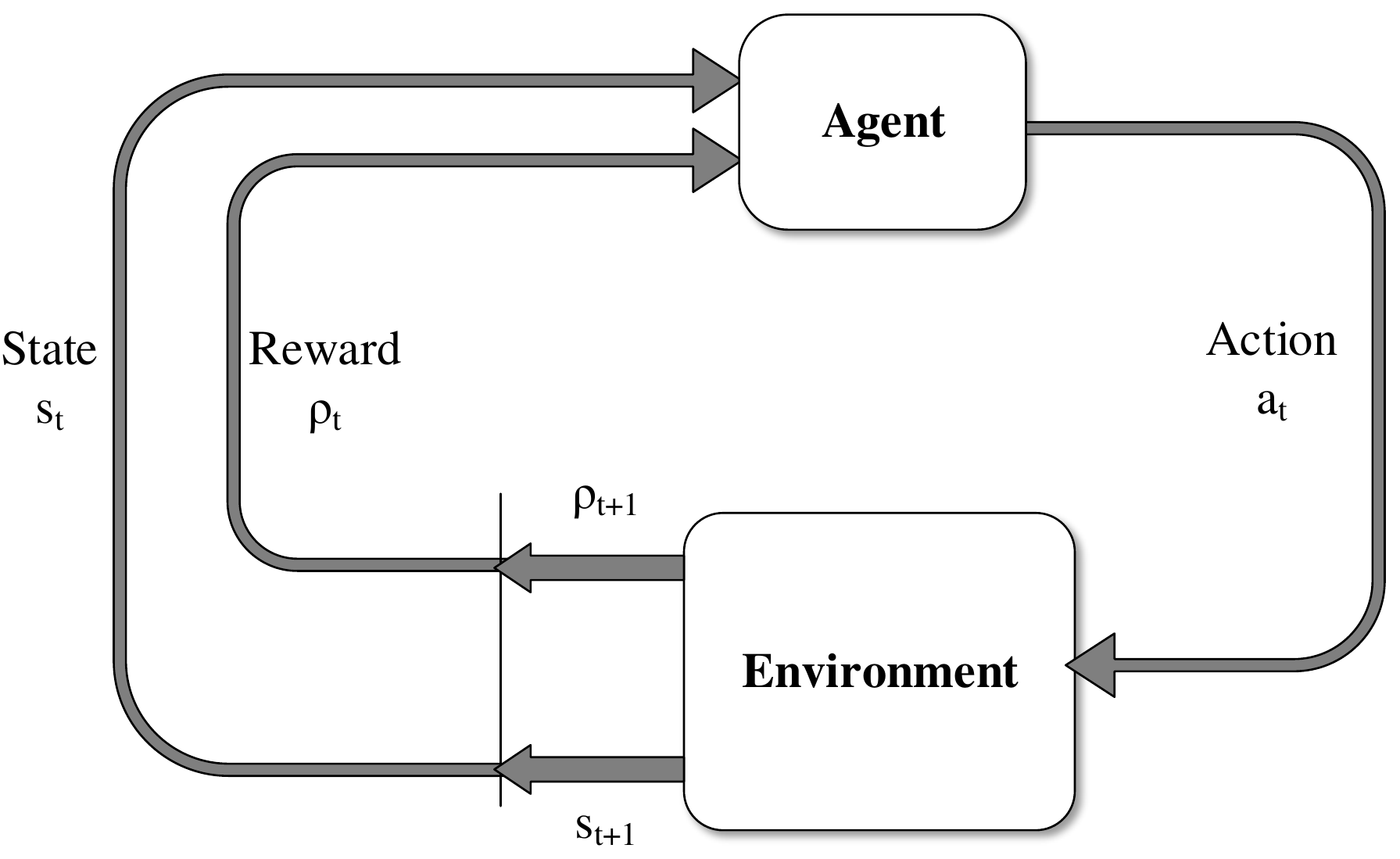}
	\caption{An agent-environment interaction~\cite{sutton2018reinforcement}.}\label{fig:agent_env}
\end{figure}	

In this section, we present our reinforcement learning-based method for available bandwidth estimation where we maximize a cumulative reward function by employing the exploration-exploitation mechanism and learn through environment observations without having a training phase. In the sequel, we start with the $\varepsilon$-greedy search algorithm and then discuss the reward function mechanism and the convergence speed of our method.
%
%
\subsection{$\varepsilon$-greedy Algorithm }
Let us consider a finite-state Markov decision process (MDP) with an agent and an environment, as shown in Fig.~\ref{fig:agent_env}. Let us further consider that there are a set of states, $\mathcal{S}$, a set of actions, $\mathcal{A}$, and a set of rewards, $\mathcal{R}$. Here, we assume that there exists a bijective function between the sets of actions and states and the set of rewards. Particularly, there exists a one-to-one correspondence between the action-state pairs and the rewards. At time $t$, the agent in state $s_{t} \in \mathcal{S}$ chooses an action $a_{t} \in \mathcal{A}(s_{t})$, and the environment returns a reward, $\rho_{t+1} \in \mathcal{R} \subset \mathbb{R}$, and changes the agent's state to $s_{t+1} \in \mathcal{S}$. Here, $\mathcal{A}(s_{t})$ refers to the set of actions that the agent chooses when it is in state $s_{t}$. Particularly, $\mathcal{A}(s_{t})$ is a subset of $\mathcal{A}$. In a stochastic environment, the reward values following an action in one state can be samples from a distribution with a mean and  variance. In this case, the reward of the action in that state can be the average of rewards received until the last time the action is chosen, and the agent is in that state. Under these conditions, given that the agent is in state $s_{t}$, the $\varepsilon$-greedy algorithm chooses with probability $1-\varepsilon$ the action $a_{t}\in\mathcal{A}(s_{t})$ that performs the best with respect to reward returns, and selects uniformly one action among the others with probability $\varepsilon$. Particularly, the algorithm guides the agent with the best action observed while exploring with probability $\varepsilon$ among the other actions to find a better action. Here, $\varepsilon$ indicates how greedy the agent is, and the optimal value of $\varepsilon$ is important especially in noisy environments because the agent needs to explore more in order to find the action that performs best on average. When $\varepsilon$ is smaller, the agent converges to a reward value slowly and stabilizes on an action. However, although it is more stable in the long run, yet there is a risk that the reward value is not the maximum reward the agent could have. On the other hand, when $\varepsilon$ is larger, it takes shorter to converge to a reward value, but there will be too much variations in the long-run even if the measurements are not very noisy. For more details, we refer interested readers to \cite{sutton2018reinforcement}. 

In our experiment, we consider that the network is stationary, i.e., the network statistics remain constant for the time interval during which we make our measurements and estimate the average available bandwidth. Therefore, we treat available bandwidth estimation as a single-state MDP multi-armed bandit problem. Herein, the set of input probe rates,  i.e., $r_{in} \in \{\delta_r, 2\delta_r, \dots, k \delta_r\} $ in our setting corresponds to the set of actions,  $\mathcal{A}$. We further define a reward parameter, which is a function of the $r_{in}$ and $r_{out}$, and reaches the maximum when the input probe rate, $r_{in}$, is equal to the available bandwidth in the network. We provide the details of the reward function in the sequel.

Following the selection of one probe rate among $k$ input rates, its associated reward is received. Because the reward values are perturbed due to noisy measurements, we rely on the corresponding average rewards after a probing rate is selected. Particularly, we calculate the action-value function $\mathcal{Q}_{t}(r_{in})$  that estimates the value for choosing $r_{in}$  at time step $t$ by calculating the average rewards received up to time $t-1$ as
\begin{equation}\label{eqn:estimated_action_value}
\mathcal{Q}_{t}(r_{in} ) = \frac{\sum_{j=1}^{t-1}{\rho}_{j}i_{j}(r_{in})}{\sum_{j=1}^{t-1}i_{j}(r_{in})},
\end{equation}
where $i_{j}(r_{in})$ is the indicator function, which is set to 1  whenever the input rate $r_{in}$ is chosen up to time $t-1$  and is 0 otherwise. Here, the $\varepsilon$-greedy algorithm at time $t$ chooses the input probe rate that has the maximum average reward up to time $t-1$ with probability $1-\varepsilon$. Specifically, the algorithm sets the input rate at time $t$, i.e., $r_{{in}_{t}}$, as
\begin{equation*}
r_{{in}_{t}} = \arg\max_{r_{in}\in\mathcal{A}}\{\mathcal{Q}_{t}(r_{in})\}.
\end{equation*}
It uniformly chooses any rate among the others with probability $\varepsilon$ and sets the input rate.
%
%
\subsection{Choice of Reward Function}
\begin{figure*}
	\centering
	\subfigure[Reward distribution]{
		\includegraphics[width=.62\columnwidth]{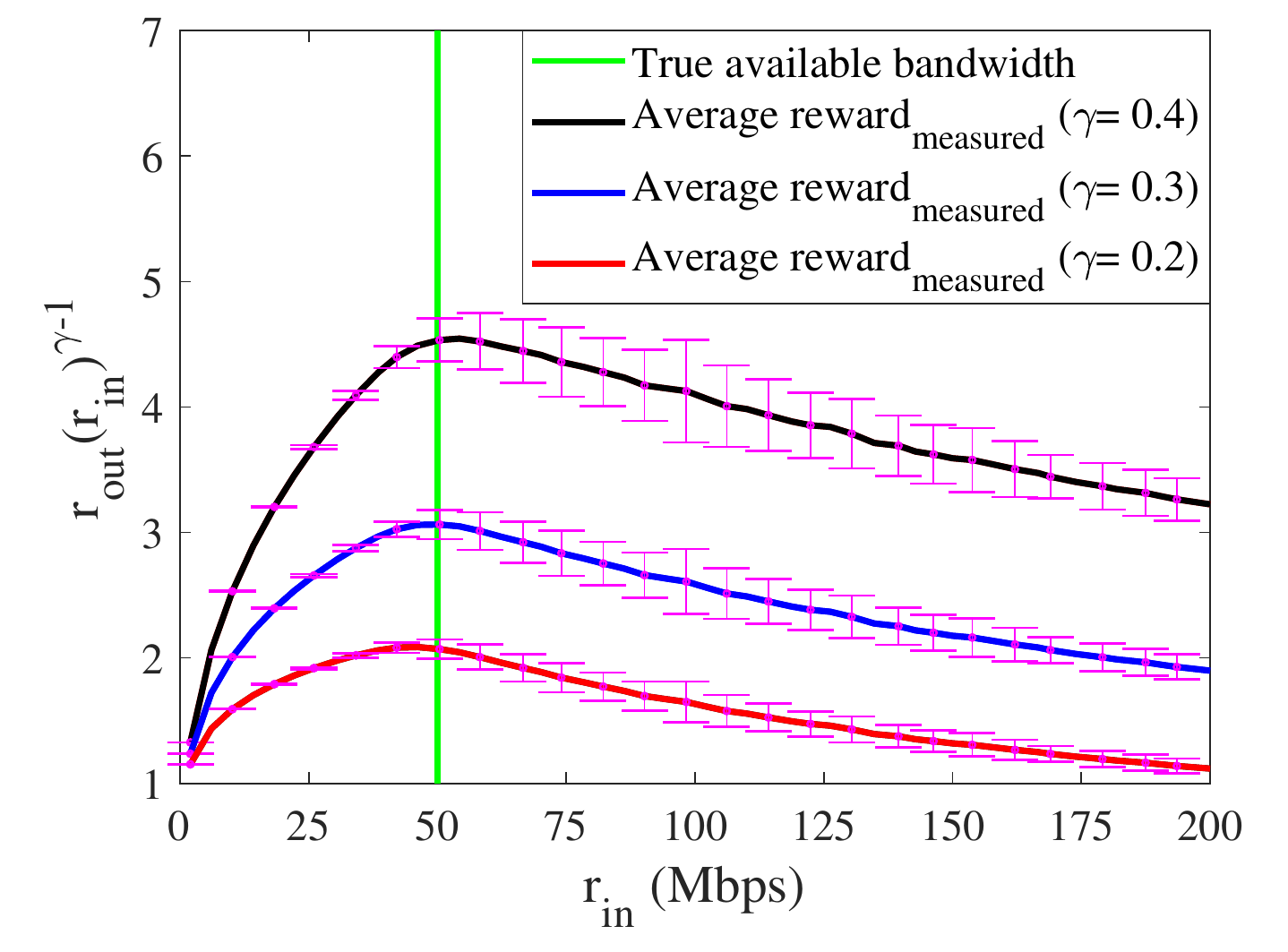}
		\label{fig:measured_reward}
	}
    \hfil
	\subfigure[ Effect of $\gamma$ on convergence speed]{
		\includegraphics[width=.62\columnwidth]{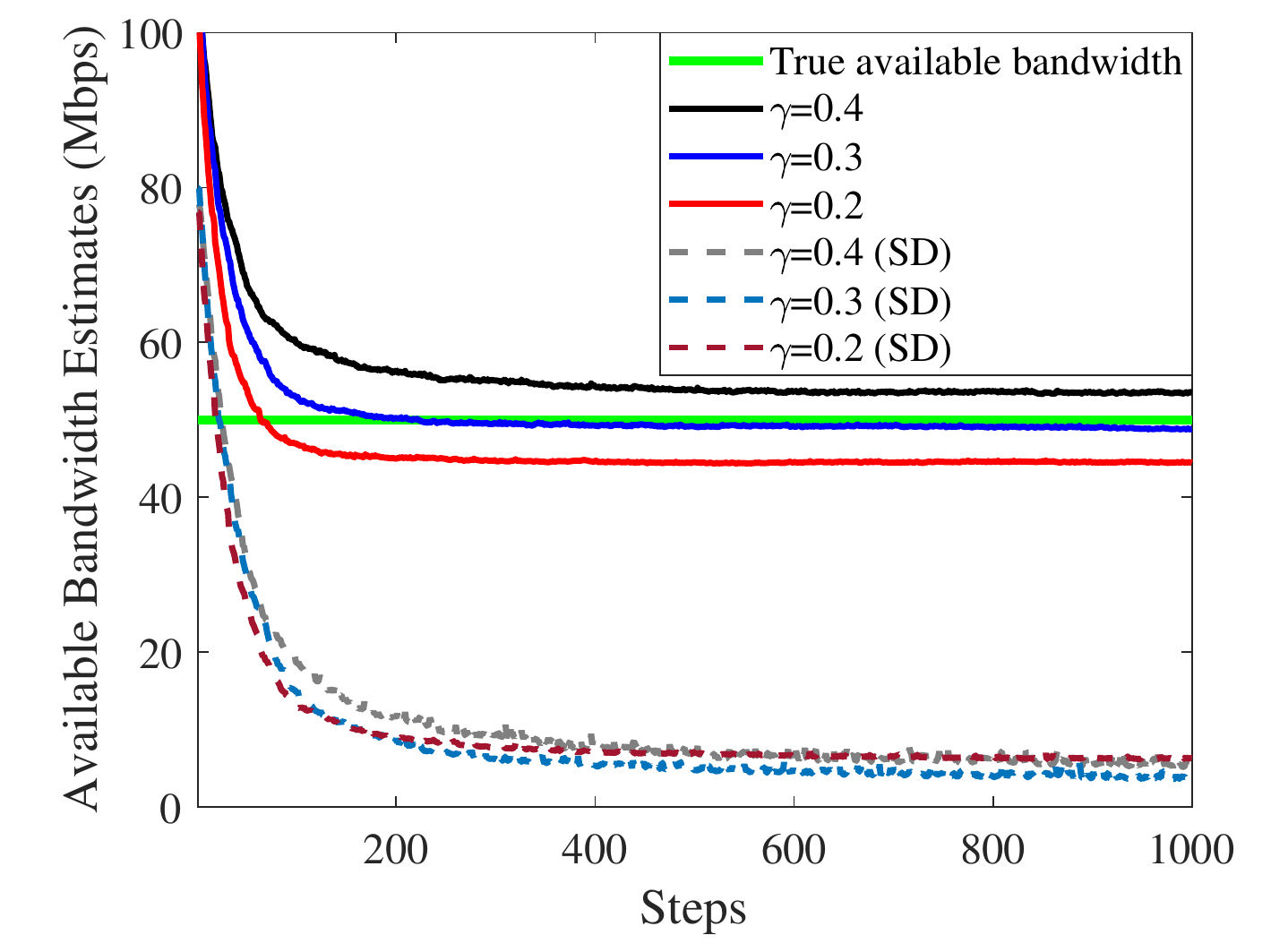}
		\label{fig:effect_gamma_conv}
	}
	\hfill
	\subfigure[Effect of $\epsilon$ on convergence speed]{
		\includegraphics[width=.62\columnwidth]{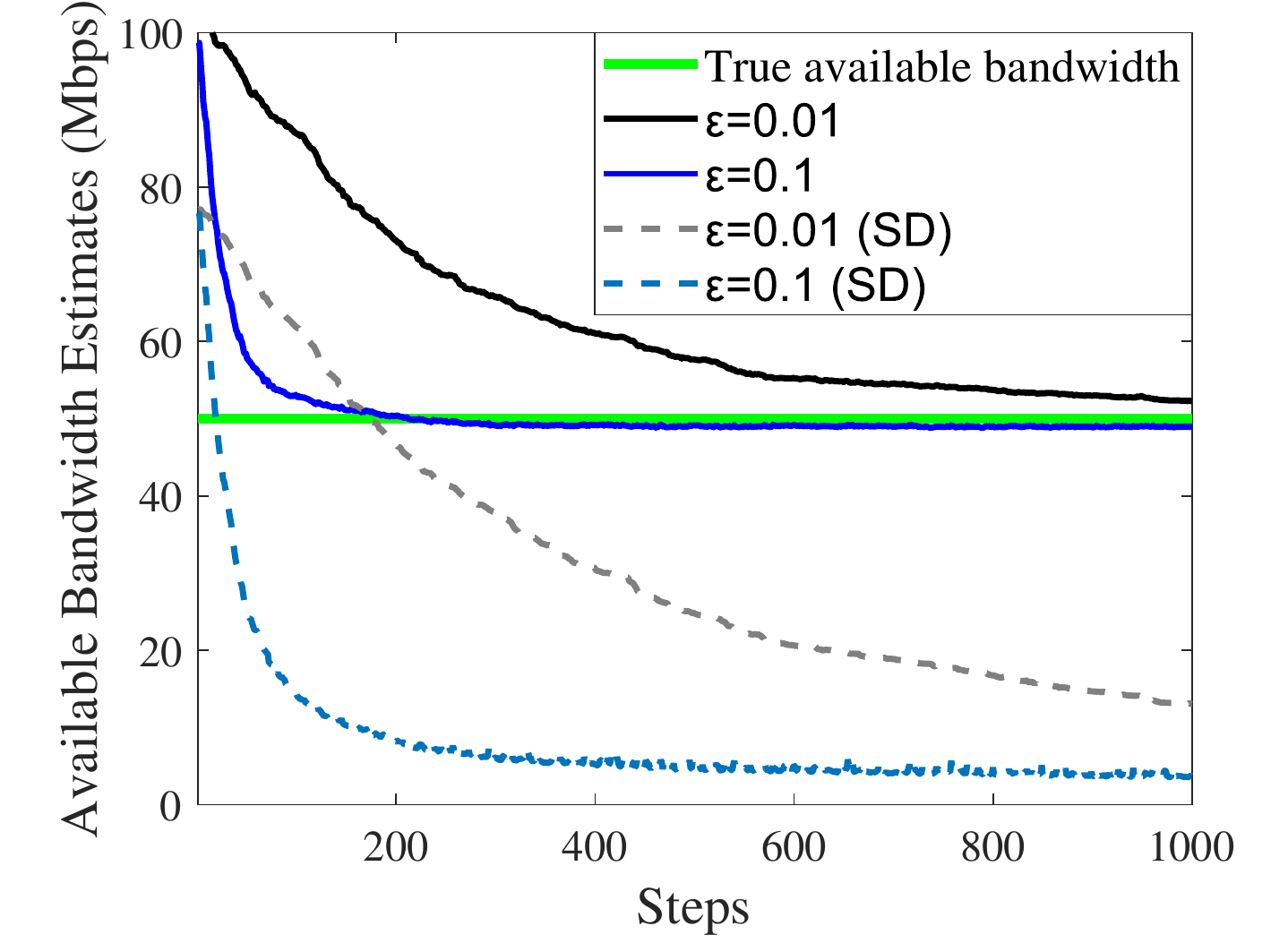}
		\label{fig:effect_epsilon_conv}
	}
	\caption{(a) Reward distribution with average measured rewards and error bars depicting their SDs, and (b) \& (c) average available bandwidth estimates and their SDs for different values of $\gamma$ and $\varepsilon$, the two parameters  that affect the convergence speed of reinforcement learning-based method.}
\end{figure*}

In real-time available bandwidth estimation, the major challenge is to define a function that produces a credible reward even in the presence of noisy measurements due to non-fluid traffic, multiple bottlenecks, and inaccurate time stamping. To combat the effect of noise, we define a reward metric which is a function of the $r_{out}$ and $r_{in}^{\gamma-1}$, where $\gamma$ is the convergence parameter satisfying $0 < \gamma < 1- \frac{\lambda}{C}$. Formally, we define the reward function as
\begin{equation}\label{eqn:reward_fun}
	{\rho} = r_{out}(r_{in})^{\gamma-1}.
\end{equation}
The reward function in~\eqref{eqn:reward_fun} reaches the maximum when $r_{in}$ is equal to the available bandwidth as long as the convergence parameter, $\gamma$, is in the aforementioned defined range. We set the exploration rate $\varepsilon=0.1$ and  show the measured reward function for convergence parameter $\gamma\in[0.2, 0.3, 0.4] $ as a function of $r_{in}$  averaged over 1000 repeated measurements in Fig.~\ref{fig:measured_reward}, taken in the network with a single tight link of capacity $C=100$~Mbps in the presence of exponential cross-traffic with an average rate of $\lambda=50$~Mbps. The cross-traffic packet size is $l=1514$ bytes, and access links are of capacity 1~Gbps. The error bars depict the standard deviation (SD) from the average reward values, which increases when probing rate reaches beyond available bandwidth, i.e.,  $r_{in} > C-\lambda$ due to building up of queues at the multiplexer. As seen in Fig.~\ref{fig:measured_reward}, the reward function is maximized when $r_{in}$ is equal to the available bandwidth, which is $50$~Mbps. We also note that with decreasing $\gamma$, the reward function also decreases, which leads to slower convergence because the impact of noise is more belligerent with decreasing reward function when differentiating the maximum reward from the others.
%
%
\subsection{Convergence Speed}
Our proposed reinforcement learning-based method is a continuous process.  Once the convergence is reached, it produces a stable value of available bandwidth estimate. However, the speed at which it  converges depends upon the choice of two parameters, i.e.,  $\gamma$ and $\varepsilon$.   
%
%
\subsubsection{Choice of $\gamma$}
In a network with unknown $C$ and $\lambda$, it is not trivial to determine $\gamma$, which depends on these unknowns by definition. To analyze the effects of $\gamma$ on the convergence speed, we plot the average available bandwidth estimates and their standard deviations (SDs) over $1000$ steps for $\gamma\in[0.2, 0.3,0.4]$,  as shown in Fig.~\ref{fig:effect_gamma_conv}. At each step, our method chooses one of the input rates among $k$ input rates following the $\varepsilon$-greedy algorithm and provides a single available bandwidth estimate.  As  the number of steps increases, every input rate is sampled enough number of times leading to the convergence of the input rate with maximum reward value to the available bandwidth. Furthermore, we use standard deviation (SD) as a metric to measure the precision of the bandwidth estimates:
\begin{equation}\label{eqn:sd}
SD=\sqrt{\frac{1}{K_r-1}\sum_{i=1}^{K_r}(\hat{A}_{i}-\bar{A}_{i})^2},
\end{equation}
where $\hat{A}$ and $\bar{A}$ are the estimated  and the true values of the available bandwidth, respectively  and $K_r$ is the number of repeated experiments over which we obtain the average available bandwidth estimates and their SDs around the true available bandwidth. We set  $K_r=1000$ and the exploration rate to $\varepsilon=0.1$ unless otherwise stated.  As seen in Fig.~\ref{fig:effect_gamma_conv}, the convergence is faster when $\gamma=0.3$, i.e., the method detects the available bandwidth after 200 steps. On the other hand, it takes more than 1000 steps on average for the method to converge to the available bandwidth when $\gamma=0.2$ and $\gamma=0.4$. We plot the graphs until 1000 steps for clarity in the comparison of different $\gamma$ values. One can run the experiment for more steps and can easily observe  that as long as the convergence parameter satisfies $0 < \gamma < 1- \frac{\lambda}{C}$, the method will converge. However, the  convergence speed  depends not only on $\gamma$ but on the  $\varepsilon$ as well.  
%
\subsubsection{Choice of $\varepsilon$}
The choice of $\varepsilon$ dictates the exploration-exploitation trade-off in reinforcement learning-based methods. Hence, in order to understand the impact of $\varepsilon$, we plot the average available bandwidth estimates and their SDs for $\varepsilon\in[0.01,0.1]$ with the convergence parameter set to $\gamma=0.3$. The larger exploration rate, $\varepsilon = 0.1$, leads the method to explore more and  find the available bandwidth faster when compared to the smaller exploration rate, $\varepsilon= 0.01$. However, although the method converges more quickly with larger $\varepsilon$, yet the method performs better with a smaller $\varepsilon$ in the long run when the noise variance is low. This is because the method with large $\varepsilon$ is able to reach an acceptable range of $r_{in}$ that involves the available bandwidth. However, it leads to more variations in the available bandwidth estimation in the long-run since it tests other values very often. On the other hand, the method reaches a smaller range of $r_{in}$ that maximizes the reward function very slowly when $\varepsilon$ is smaller, but the variation around the available bandwidth is much smaller in the long-run.

%
\section{Model-based Reference Implementation}\label{sec:referenceimplementation}
We compare our method with the piece-wise linear network model-based direct probing technique that employs a Kalman filter, which is provided in~\cite{ekelin2006real}. While available bandwidth estimation tools differ significantly regarding the selection and the amount of probe traffic. We test both techniques using the same database for the sets of input rates and output rates in order to provide a solid reference point. In the following section, we briefly describe the direct probing technique. For more information, we refer interested readers to \cite{ekelin2006real}.
%
%
\subsection{Direct probing}
In order to implement the direct probing technique in our testbed, we combine the active probing  with a Kalman filter. Unlike in~\cite{ekelin2006real}, to increase the convergence speed of the filter, we use a multi-rate probe stream of the $k$ packet trains that correspond to $k$ input rates $r_{in}$ as in~\cite{sedighizad2012mr} to probe the network path with several rates in each experiment. We define the inter-packet strain as~\cite{ekelin2006real}
\begin{equation}
\xi= \frac{r_{in}}{r_{out}}-1 \quad  \text{for } r_{\mathrm{in}} > C - \lambda.\! \\
\label{eqn:prm3}  
\end{equation}
After inserting $\xi$ into~\eqref{eqn:prm} for  $r_{in} > C - \lambda$, we rewrite~\eqref{eqn:prm} as
\begin{equation}\label{eqn:prm_sub}
\xi  = r_{\mathrm{in}}\frac{1}{C}+ \frac{\lambda-C}{C}.
\end{equation}
By defining $\alpha = \frac{1}{C}$ and  $\beta = \frac{\lambda-C}{C}$ we obtain the packet strain parameter as
\begin{equation}\label{eqn:strain}
\xi =  \begin{cases}
0,       & \quad \text{if } r_{in} \leq C - \lambda,\\
\alpha r_{in}+\beta, & \quad \text{if } r_{in} > C - \lambda. \\
\end{cases}
\end{equation}
Following the assumptions of the fluid-flow model, we can see that the expected value of $\xi$ is  zero in the absence of congestion, and it grows proportional to the probe traffic when the probing rate exceeds the available bandwidth. As seen in~\eqref{eqn:strain}, the model is piece-wise linear due to the sharp bend at $r_{in}=C-\lambda$ which inhibits the direct application of the Kalman filter. In order to overcome the problem, we feed only the measurements that satisfy $r_{in}>\hat{A}$ to the filter, where $\hat{A}$ is the recent estimate of the available bandwidth. Since the direct probing technique seeks to estimate the upward line segment of the rate response curve which is determined by two parameters $C$ and $\lambda$, we can express the state of the system with a state vector containing two unknown parameters as
\begin{equation}
{x}_{t} =\Big[\begin{matrix}\alpha_{t}\\\beta_{t}\end{matrix} \Big].
\label{eqn:state}  
\end{equation}
Assuming that the network statistics remain constant during the observation period $t$, the transition matrix $\mathsf{A}$ becomes an identity matrix. Hence, we define the system state as~\cite{ekelin2006real}
\begin{equation}\label{eqn:system_state}
{x}_{t}= {x}_{t-1}+ {w}_{t-1},
\end{equation}
where ${w}_{t-1}$ is the process noise and denotes the deviations from the fluid-flow model. 

We define the measurement model as
\begin{equation}\label{eqn:system_measurement}
z_{t} = H_{t}x_{t} + v_{t},
\end{equation}
where $z_{t}$ is a $k\times1$ dimensional vector of measured strains as
\begin{equation}\label{eqn:ind_strain}
	{z}_{t} =\Big[\begin{matrix} z_{t}^{1} , z_{t}^{2} ,..,z_{t}^{k}
	\end{matrix} \Big]^{T},
\end{equation}
where $\{\}^{T}$ is the transpose operator. The $k$ packet trains corresponding to $k$ different rates, increase the variance of measurement noise of a probe stream. In order to combat this effect, we compute the strain in~\eqref{eqn:ind_strain} and  the corresponding measurement noise $v_{t}$  with covariance $R = \mathbb{E}\{v_{t}v_{t}^{T}\}$, which defines the precision of the strain measurement and is crucial for the tracking ability of the Kalman filter, for each packet train individually.   Similarly, the observation matrix $H_{t}$  consisting of different  input rates, i.e., $r_{in}$ corresponding to $k$  packet trains is defined as
\begin{equation*}
{H}_{t} =\Bigg[\begin{matrix}r_{in_t}^{1} \quad 1\\....\\r_{in_t}^{k} \quad 1\end{matrix} \Bigg].
\end{equation*}

Furthermore, the $2\times2$ covariance matrix of the process noise, $Q=\mathbb{E}\{w_{t}w_{t}^{T}\}$, describes the deviation of the system from the fluid-flow model, and it is treated as an adjustable parameter as in~\cite{ekelin2006real}. $Q$, being a symmetric matrix, provides three degrees of freedom for tuning; however, we use it in a simple form as $Q= \Lambda I$, where $I$ is the $2\times2$ identity matrix. We choose $\Lambda=10^{-2}$ in our settings because it allows faster convergence and less variations in available bandwidth estimates.
%
%
\section{Experimental Evaluation}\label{sec:experimental_evaluation}
\begin{figure*}
	\centering
	\subfigure[$K_r(1)$]{
		\includegraphics[width=0.62\columnwidth]{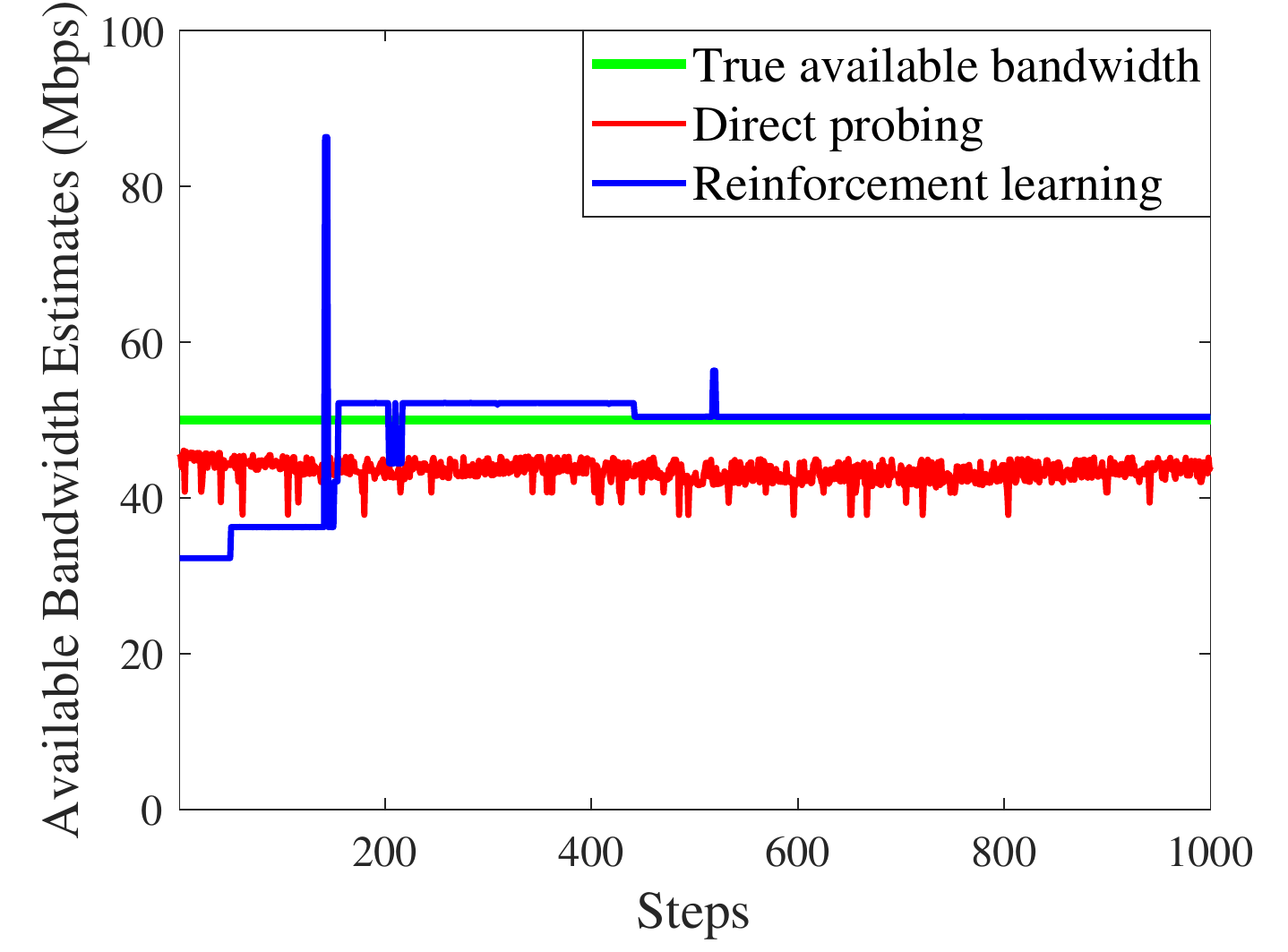}
		\label{fig:kr1}
	}
	\hfill
	\subfigure[$K_r(2)$]{
		\includegraphics[width=0.62\columnwidth]{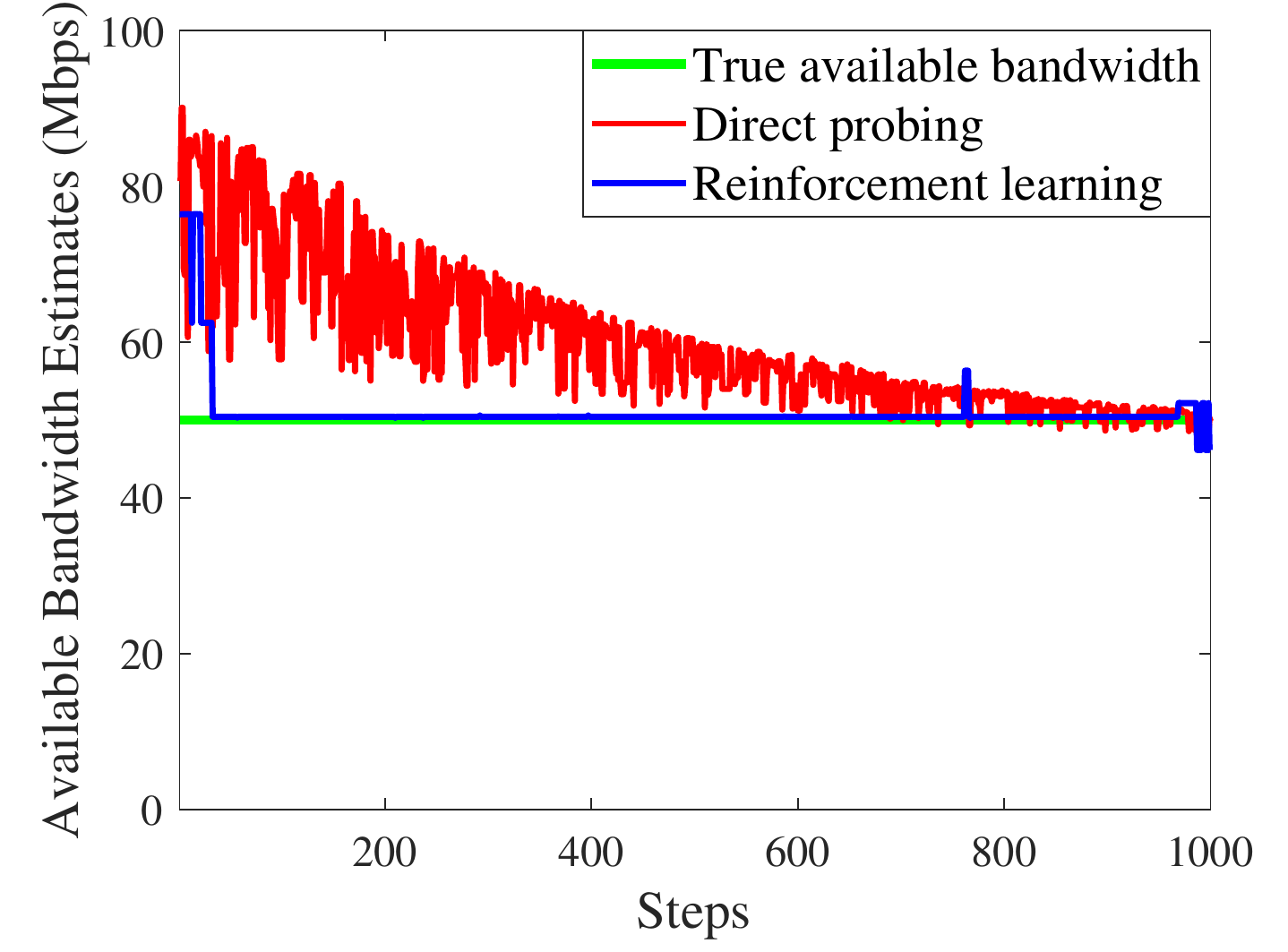}
		\label{fig:kr2}
	}
	\hfill
	\subfigure[$K_r(3)$]{
		\includegraphics[width=0.62\columnwidth]{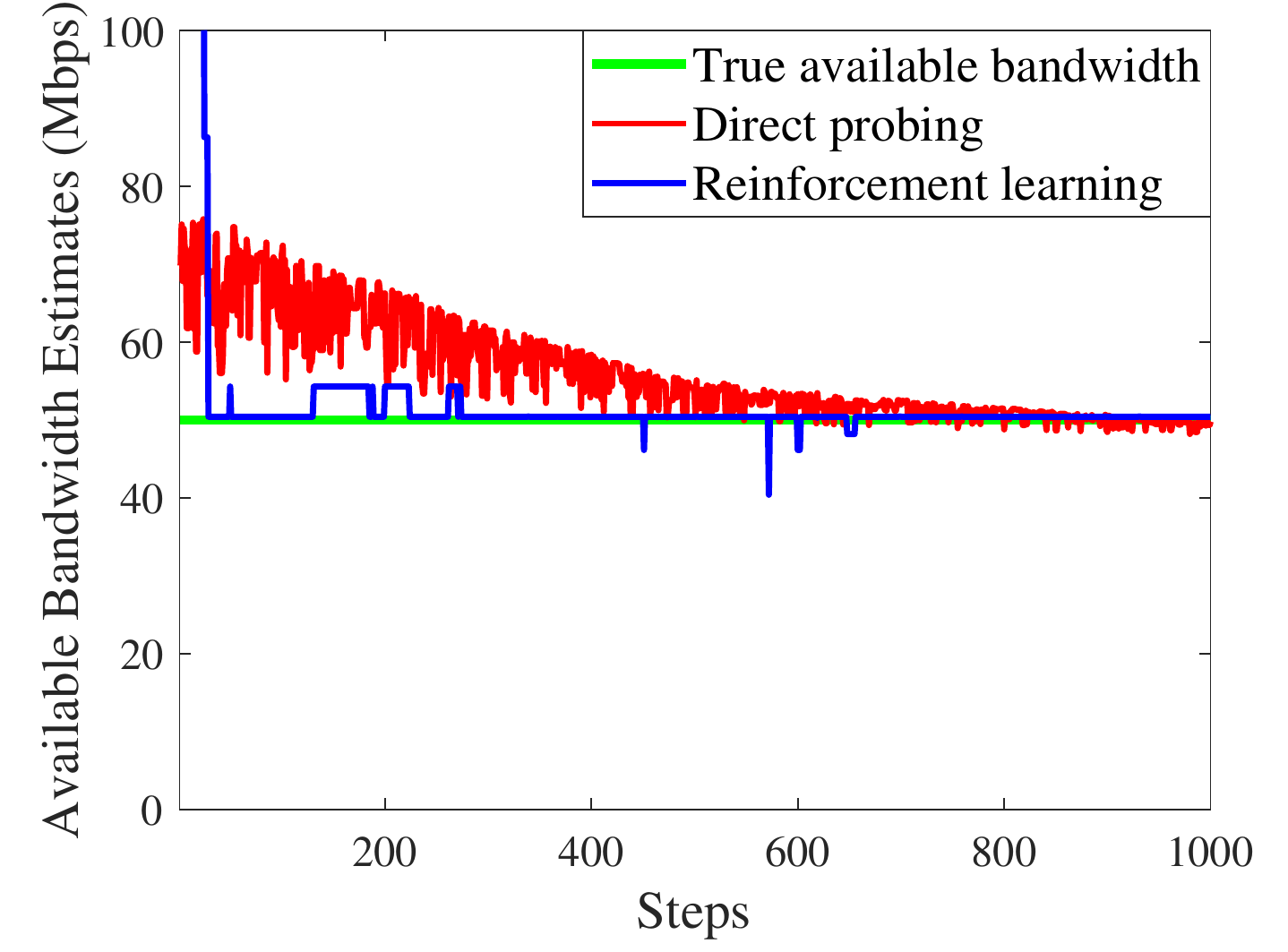}
		\label{fig:kr3}
	}
	\caption{Available bandwidth estimates for randomly selected first three repeated experiments. }
\end{figure*}
 
We evaluate the performance of our technique and compare it with the performance of the direct probing technique in a controlled network testbed described in Section~\ref{sec:exp_setup}. We use the same setting we have in Section \ref{sec:reinforcement_based} unless otherwise stated. In Fig.~\ref{fig:kr1}, Fig.~\ref{fig:kr2}, Fig.~\ref{fig:kr3}, we show randomly selected the first three repeated experiments and available bandwidth estimation results employing the direct probing technique and the reinforcement learning-based method. As seen in the randomly selected experiments, our method outperforms the other method. However,  in order to have a better view from a statistical perspective, we perform  $K_r=1000$ experiments and compare the average estimation performances and their SDs around the actual available bandwidth values in the sequel. 
%
%
\subsubsection{Cross traffic Burstiness}
\begin{figure*}
	\centering
	\subfigure[Constant Bit Rate]{
		\includegraphics[width=0.62\columnwidth]{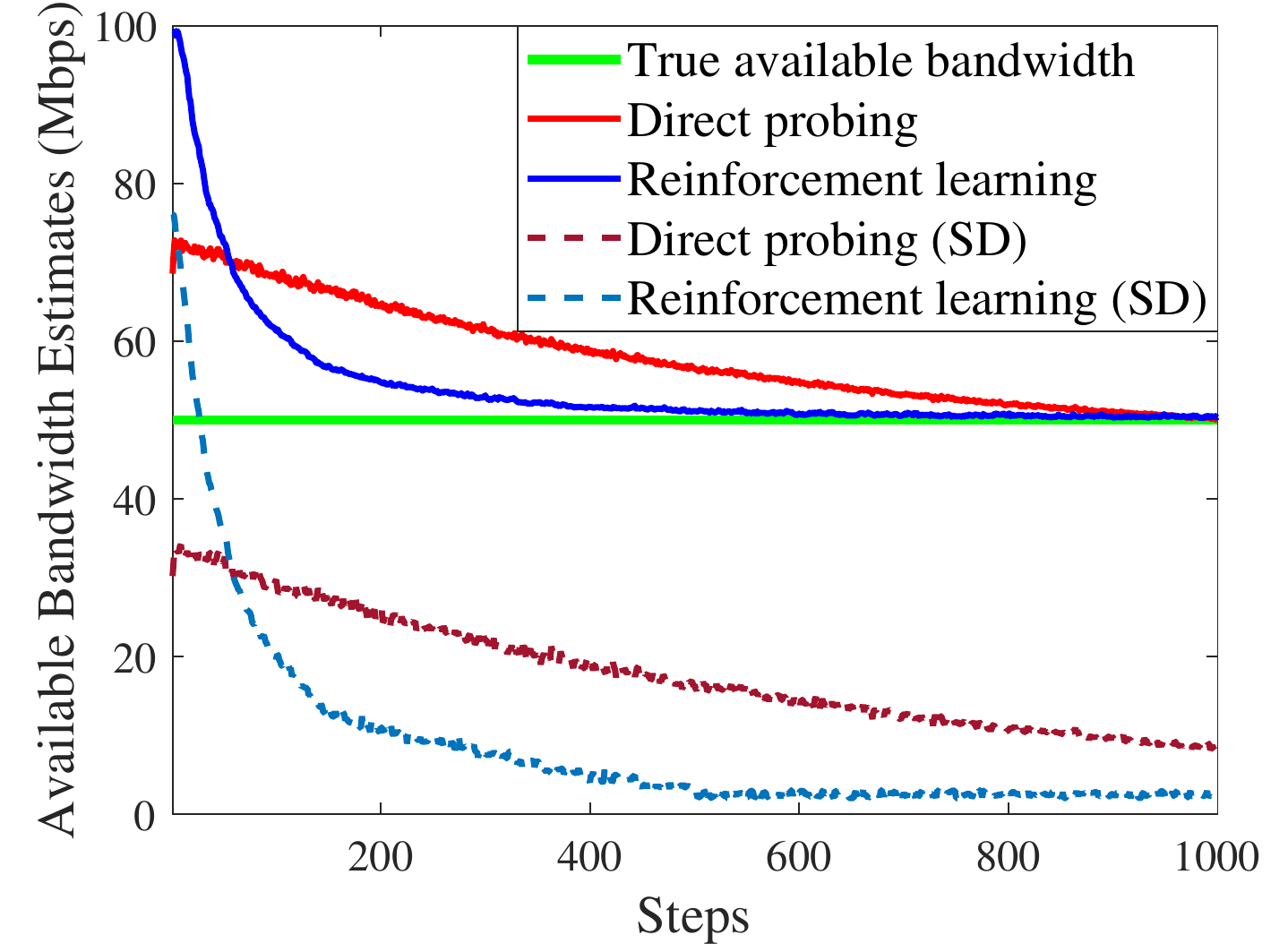}
		\label{fig:cbr}
	}
	\hfill
	\subfigure[Exponential]{
		\includegraphics[width=0.62\columnwidth]{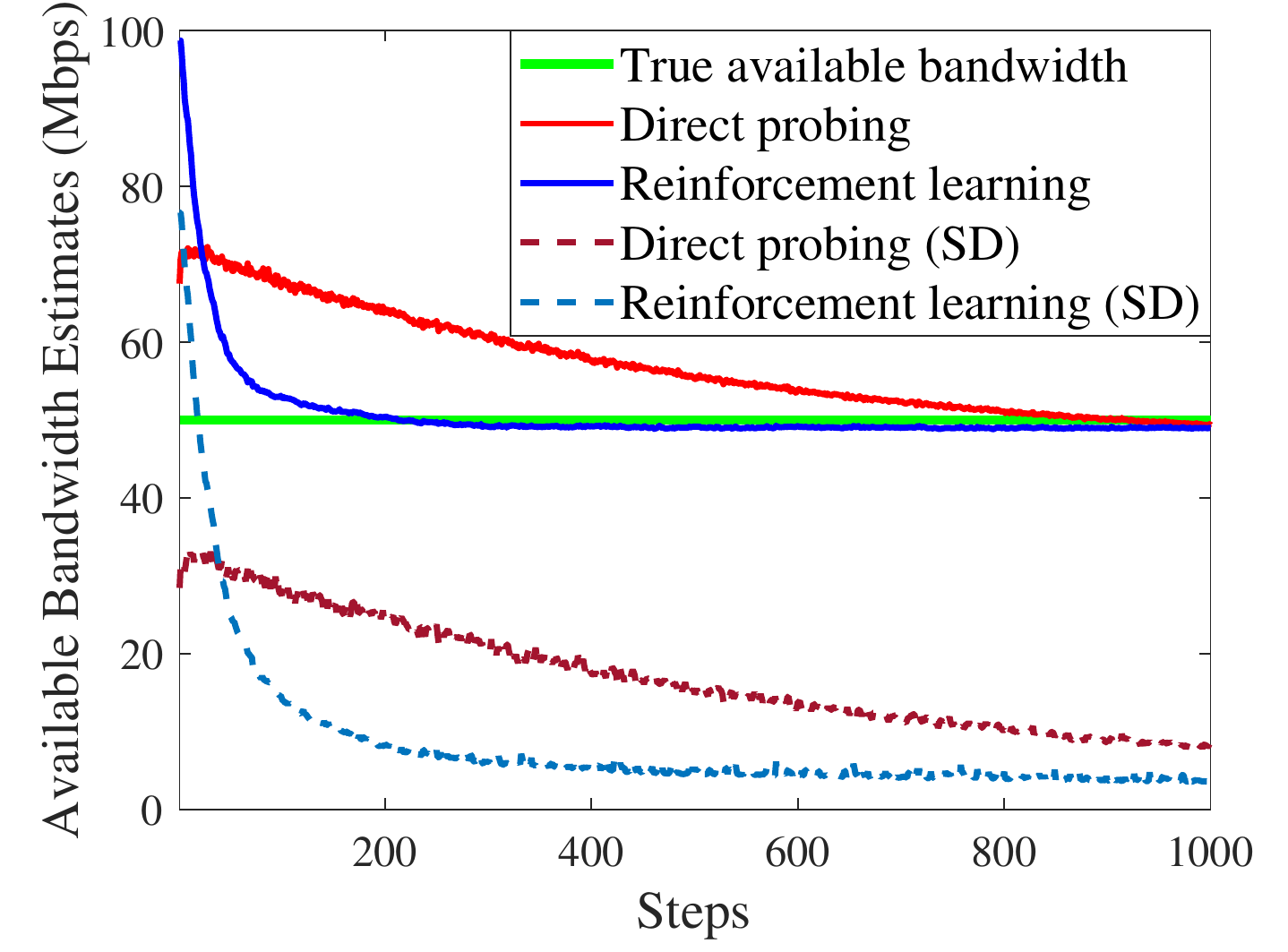}
		\label{fig:exponential}
	}
	\hfill
	\subfigure[Pareto]{
		\includegraphics[width=0.62\columnwidth]{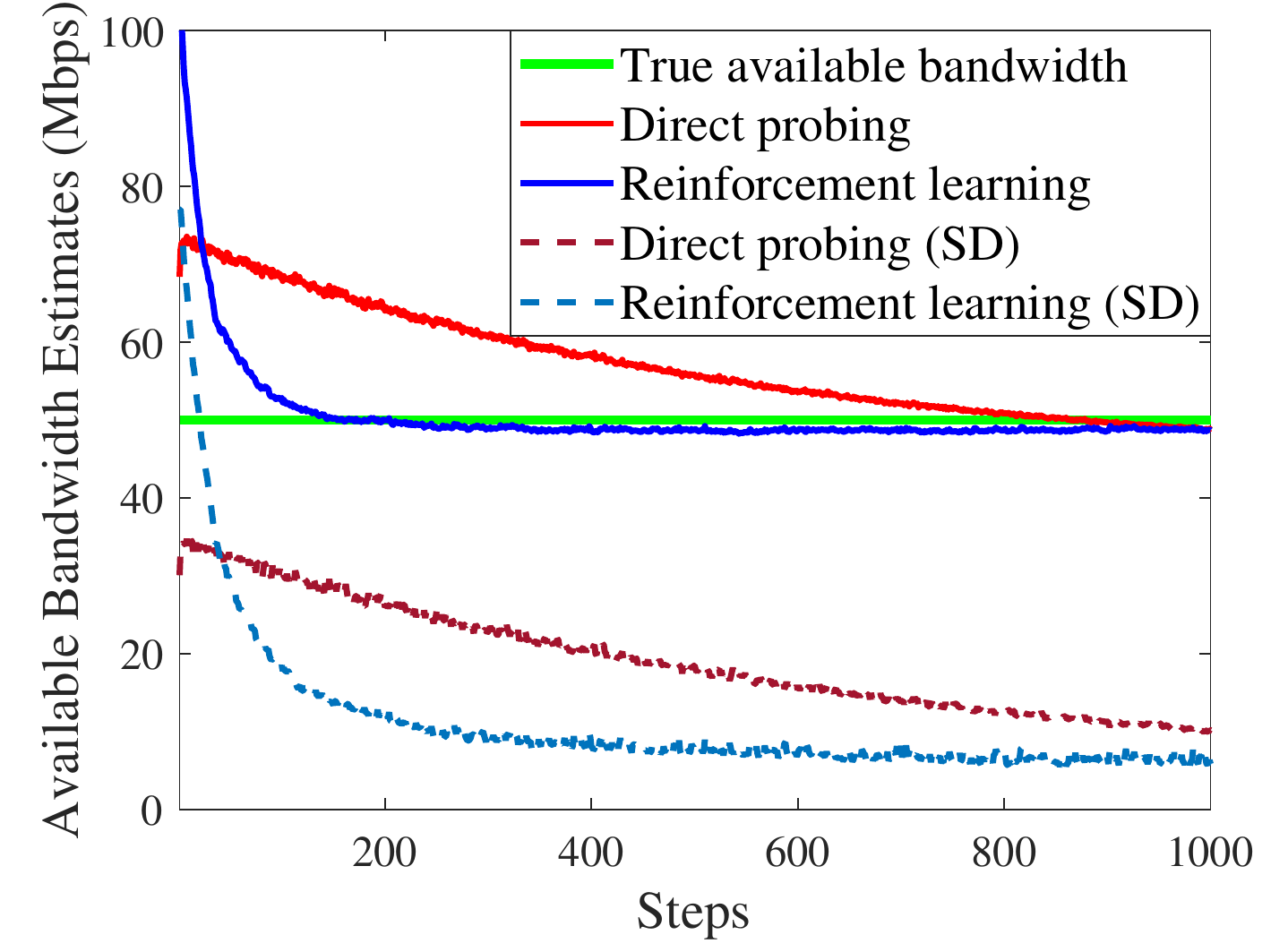}
		\label{fig:pareto}
	}
	\caption {Average available bandwidth estimates and their SDs  for different types of cross-traffic burstiness. }
\end{figure*}

\begin{figure*}
	\centering
	\subfigure[$\lambda=25~Mbps$]{
		\includegraphics[width=0.62\columnwidth]{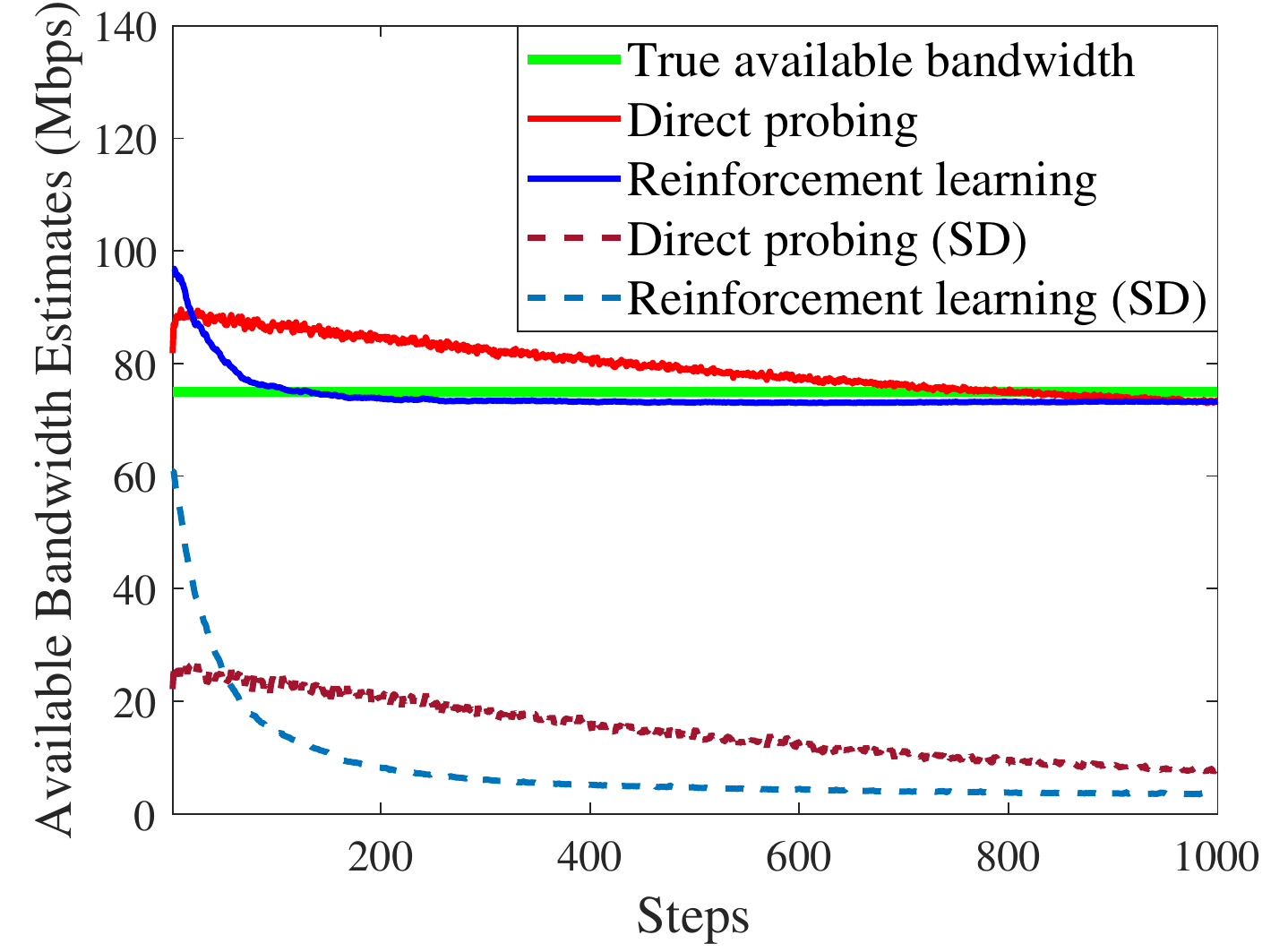}
		\label{fig:25}
	}
	\hfill
	\subfigure[$\lambda=50~Mbps$]{
		\includegraphics[width=0.62\columnwidth]{./fig/et_50_bart_reinf_sd_K_1000_mean_cr}
		\label{fig:50}
	}
	\hfill
	\subfigure[$\lambda=75~Mbps$]{
		\includegraphics[width=0.62\columnwidth]{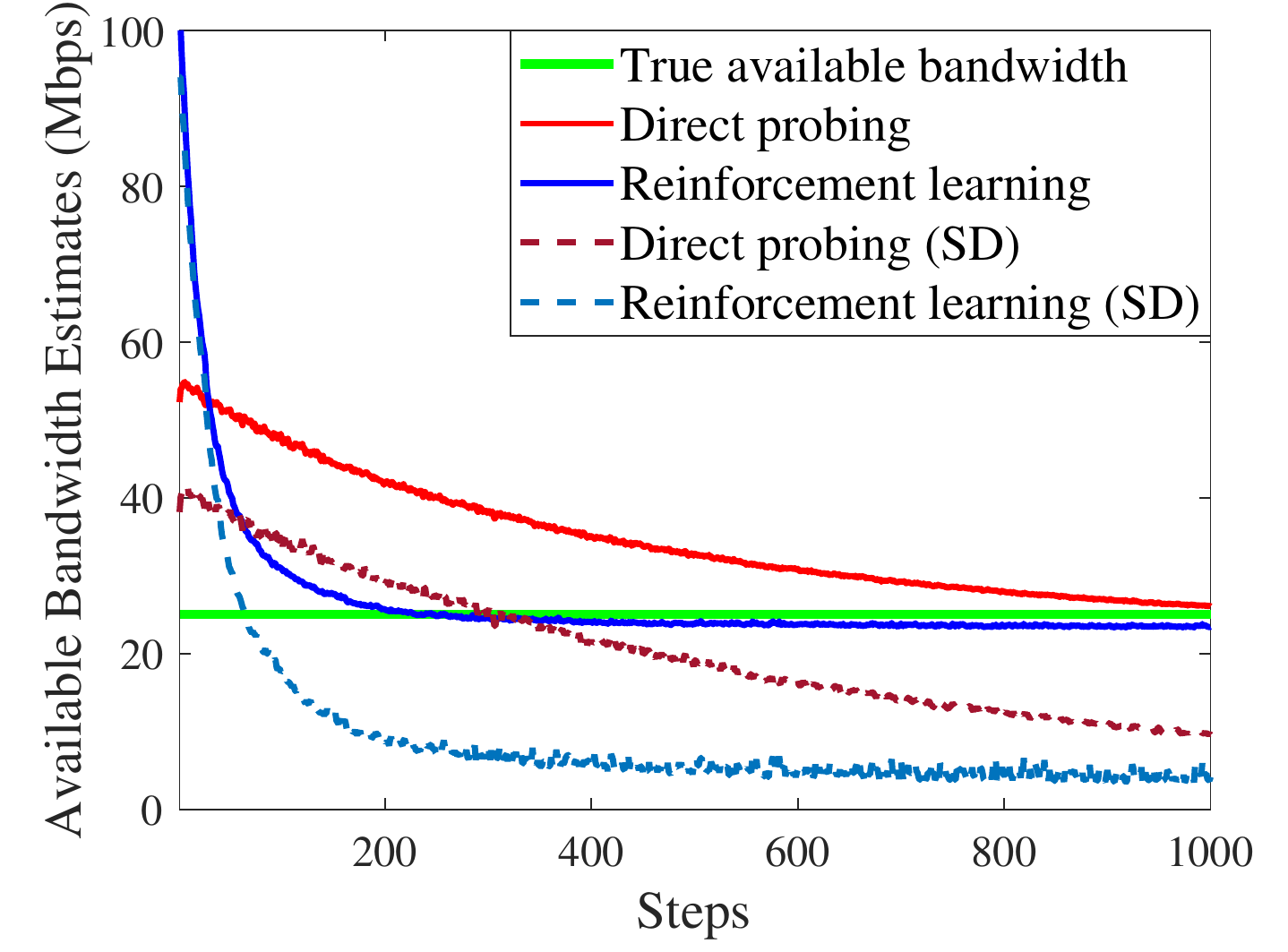}
		\label{fig:75}
	}
	\caption{Average available bandwidth estimates and their SDs for different  exponential cross-traffic rates $\lambda \in {25,50,75}$~Mbps. }
\end{figure*}

In order to evaluate how our method performs in the presence of cross-traffic with an unknown burstiness, we consider three types of cross-traffic:
\begin{enumerate}
	\item No burstiness with constant bit rate,
	\item Moderate burstiness due to exponential packet inter-arrival times,
	\item Heavy burstiness due to Pareto inter-arrival times with infinite variance, defined by a shape parameter $\alpha = 1.5$.
\end{enumerate}
Recall that the burstiness of the cross-traffic can cause queueing at the tight link even if the probe rate is below the available bandwidth, i.e., if $r_{in} < C-\lambda$, which leads to deviations from the ideal rate response curve. We show these deviations in Fig.~\ref{fig:rate_res_dev}, and we observe the maximum deviation when $r_{in}=C-\lambda$. Moreover, the strong deviation blurs the bend that marks the available bandwidth and causes an estimation bias. As seen in Fig.~\ref{fig:cbr}, Fig.~\ref{fig:exponential} and Fig.~\ref{fig:pareto}, the bandwidth estimates of the reinforcement learning-based method are more accurate with low SDs when compared to the direct probing technique. We can see the significant improvement in the convergence speed when we employ the reinforcement learning-based method  irrespective of the cross-traffic burstiness. Particularly, the reinforcement learning-based method is robust to the deviations from the fluid-flow model. 
%
%

\subsubsection{Cross Traffic Intensity}
To evaluate the impacts of cross-traffic intensity on available bandwidth estimation, we deploy exponential cross-traffic with average rates $\lambda = 25$, $50$, and $75$ Mbps, and depict the average of the available bandwidth estimates and their SDs around the true available bandwidth in Fig.~\ref{fig:25}, Fig.~\ref{fig:50} and Fig.~\ref{fig:75}, respectively. While the SDs increase in the direct probing technique with the increasing cross-traffic, the SDs in the reinforcement learning-based method remains almost unchanged in all cases. Moreover, the reinforcement learning-based method converges to the available bandwidth faster than the direct probing technique does.

%
\subsubsection{Multiple Tight Links}

\begin{figure}
	\centering
	\subfigure[1-hop]{
		\includegraphics[width=0.56\columnwidth]{./fig/et_50_bart_reinf_sd_K_1000_mean_cr}
		\label{fig:1hop}
	}
	\hfill
	\subfigure[2-hop]{
		\includegraphics[width=0.56\columnwidth]{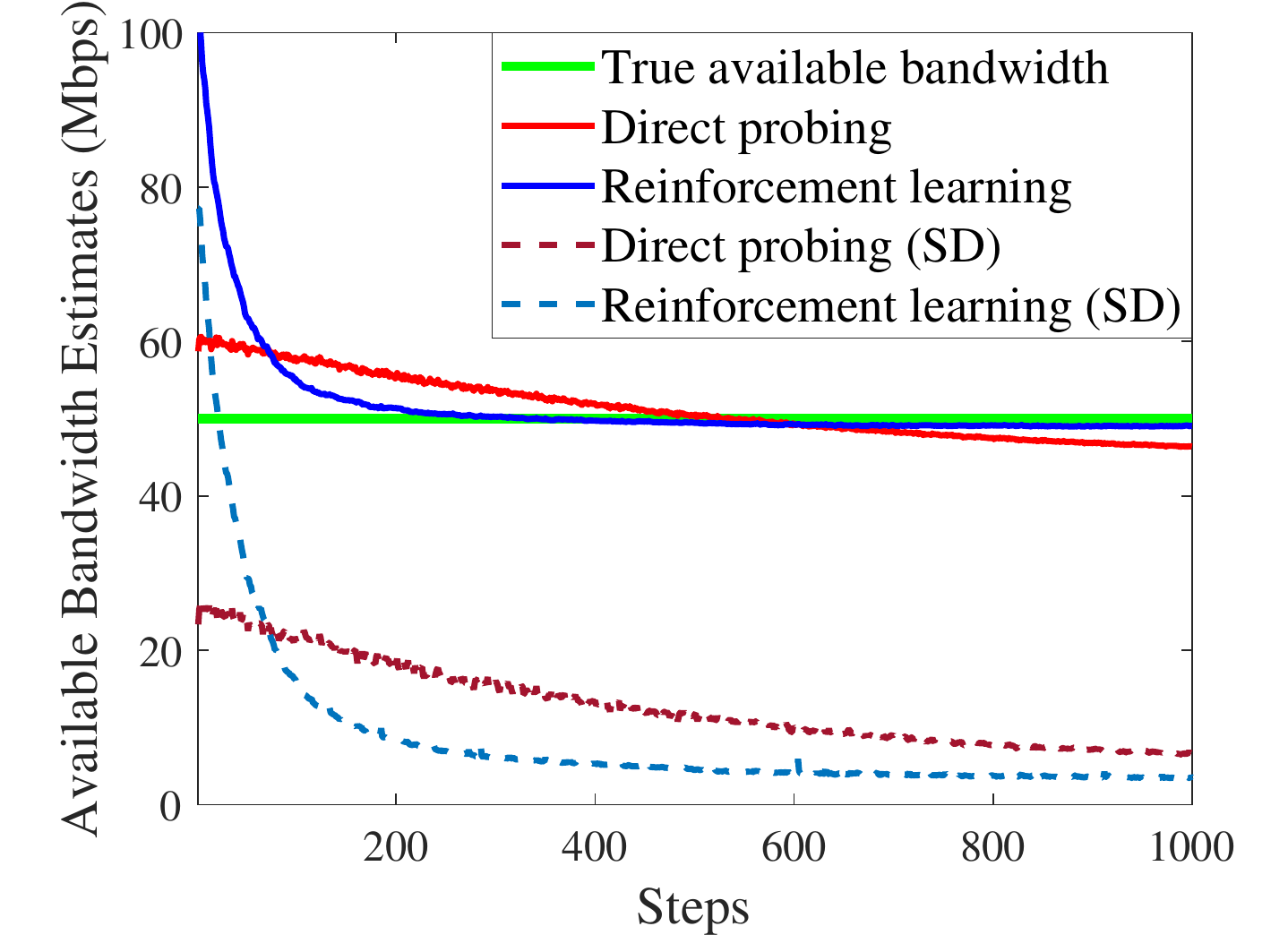}
		\label{fig:2hop}
	}
	\hfill
	\subfigure[3-hop]{
		\includegraphics[width=0.56\columnwidth]{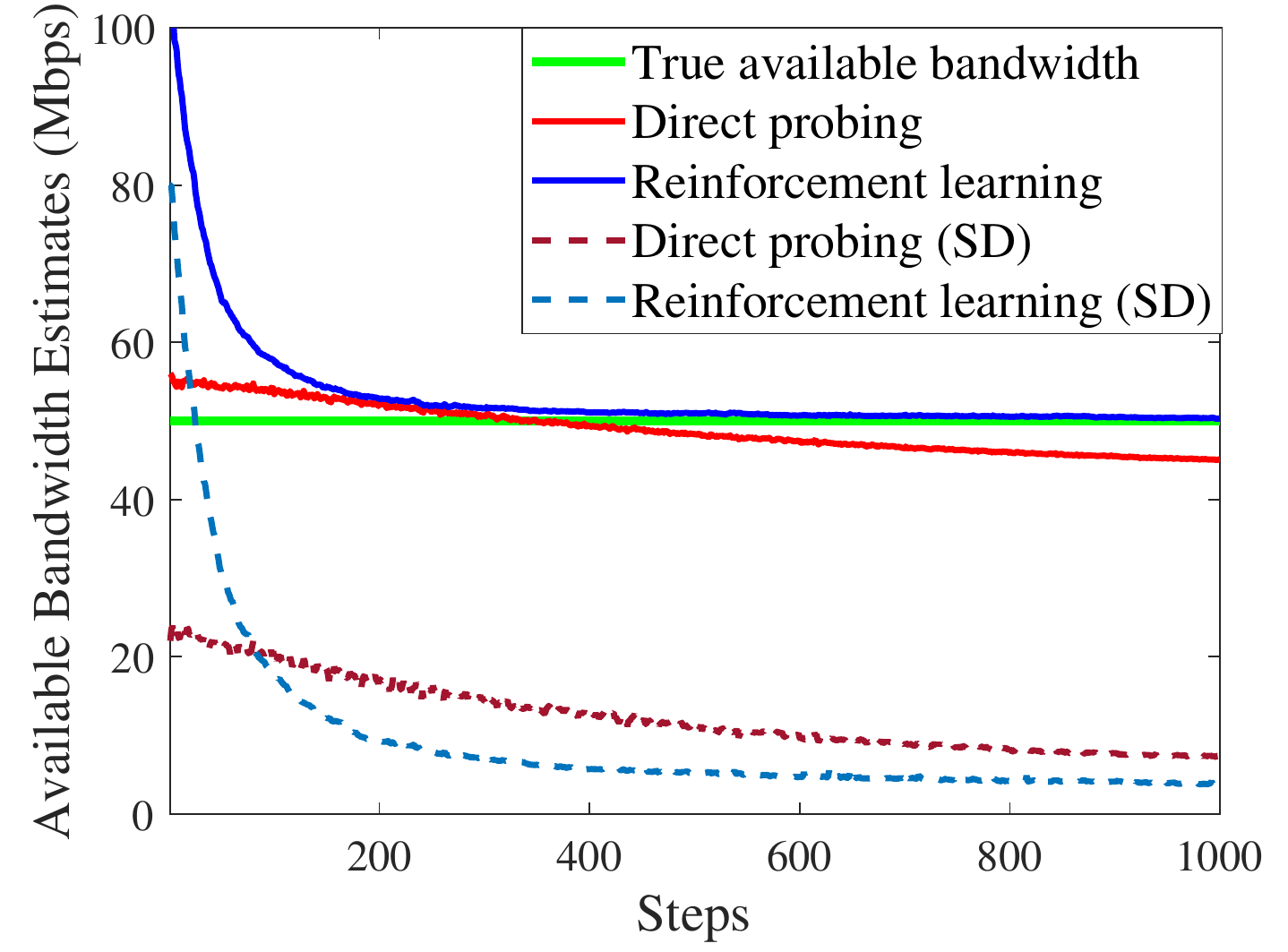}
		\label{fig:3hop}
	}
	\hfill
	\subfigure[4-hop]{
		\includegraphics[width=0.56\columnwidth]{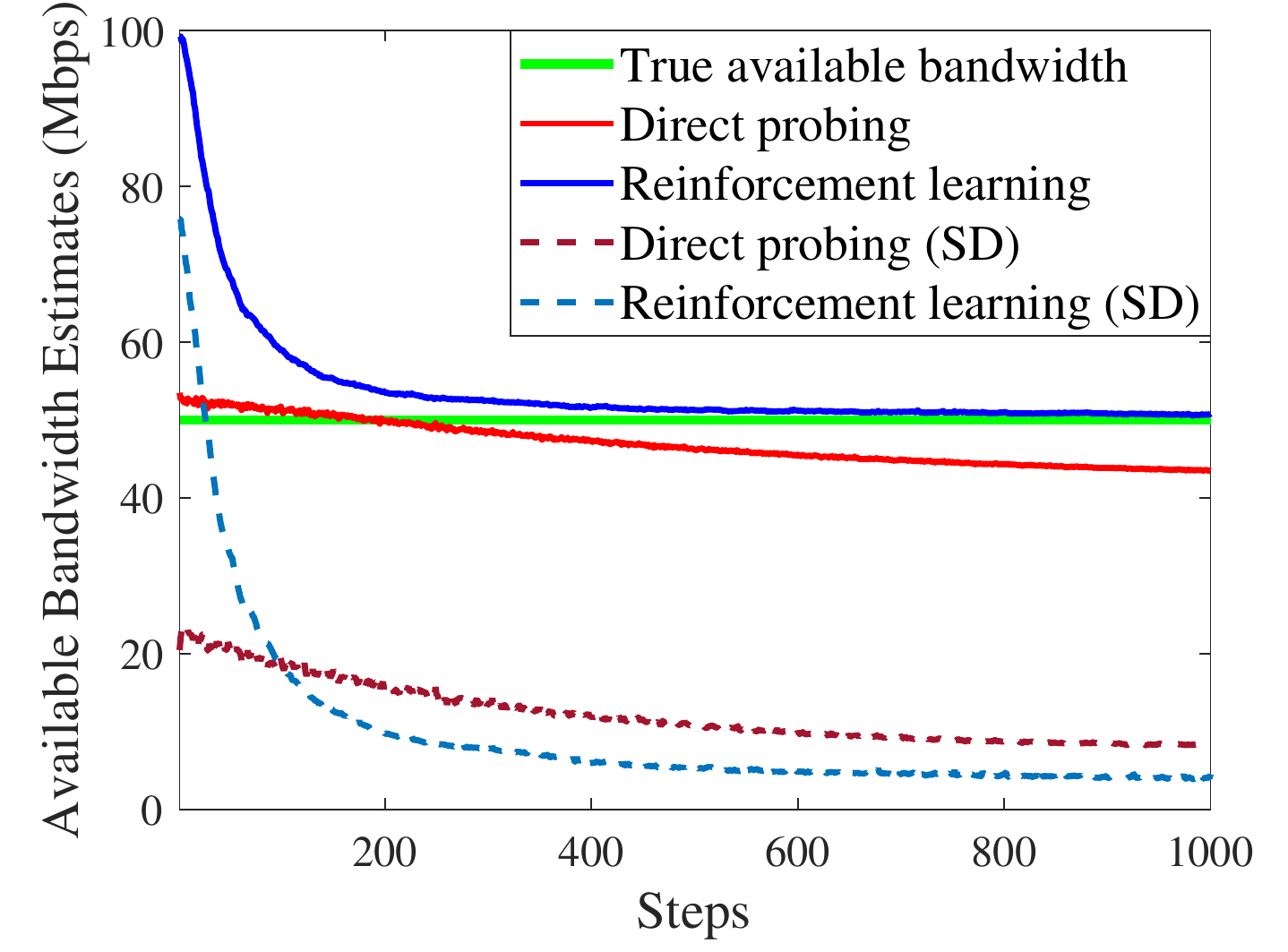}
		\label{fig:4hop}
	}
	\caption{Average available bandwidth estimates and their SDs for multiple tight links with capacity $C=100$~Mbps in the presence of single hop-persistent exponential cross-traffic with an average rate $\lambda=50$~Mbps.}
	\label{fig:multi_link}
\end{figure}

We extend our testbed from the single-hop network to the multi-hop network, as shown in Fig.~\ref{fig:topo}, to test the reinforcement learning-based method in multiple tight links. While traversing the entire network path with the tight link capacity $C=100$~Mbps and the access links with capacity 1~Gbps, the path-persistent probe streams experience single hop-persistent cross-traffic with exponential packet inter-arrival times  and average rate $\lambda=50$~Mbps. We show in Fig.~\ref{fig:multi_link} that the reinforcement learning-based method provides more accurate available bandwidth estimates, whereas the other method fails to converge to the available bandwidth. This can be explained by the fact that in the case of multiple tight links, the probe stream has a constant rate, $r_{\mathrm{in}}$, with a defined input gap, $g_{\mathrm{in}}$, only at the first link. In the following links, the input gaps have a random structure as they are the output gaps from the preceding links~\cite{jain2004fallacies, liu2007queueing, luebben2014servicecurve}. For the direct method, the inter-packet strain, $\xi$, does not grow linearly with the cross-traffic in multi-hop networks~\cite{bergfeldt2010available}, which causes the underestimation of the available bandwidth. 
%
%
\subsubsection{Tight Link but not Bottleneck Link}

\begin{figure}
	\centering
	\includegraphics[width=1\columnwidth]{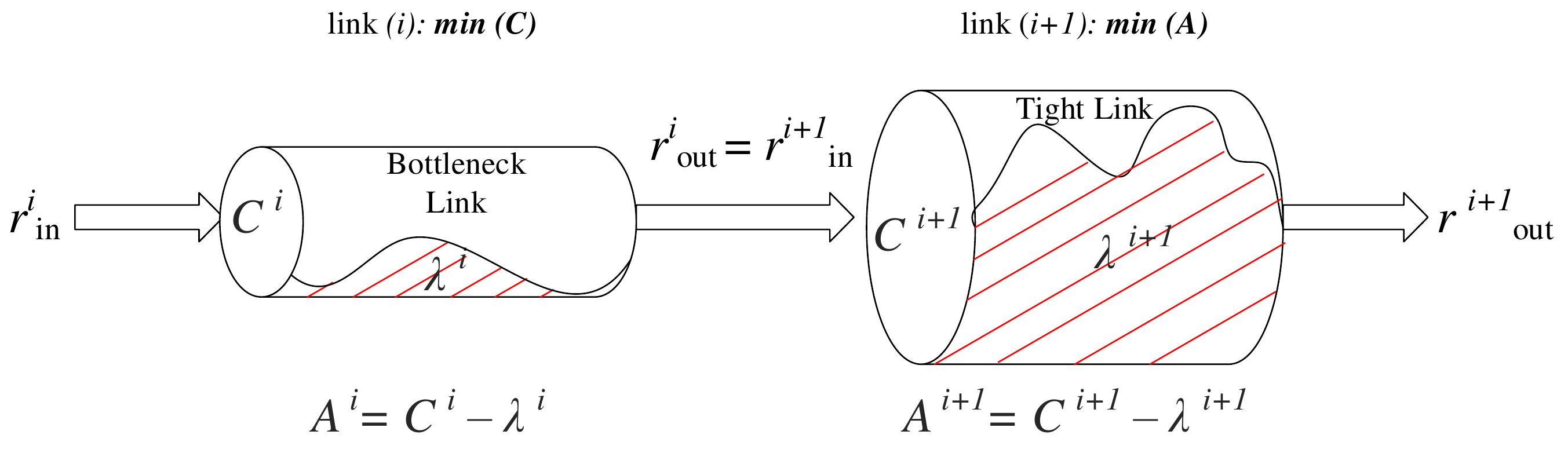}
	\caption{A two-hop network where the tight link differs from the bottleneck link~\cite{khanguraml}.}
	\label{fig:tight_bottle}
\end{figure}	

\begin{figure*}
	\centering	
	\subfigure[Rate response curves~\cite{khanguraml}]{
		\includegraphics[width=0.62\columnwidth]{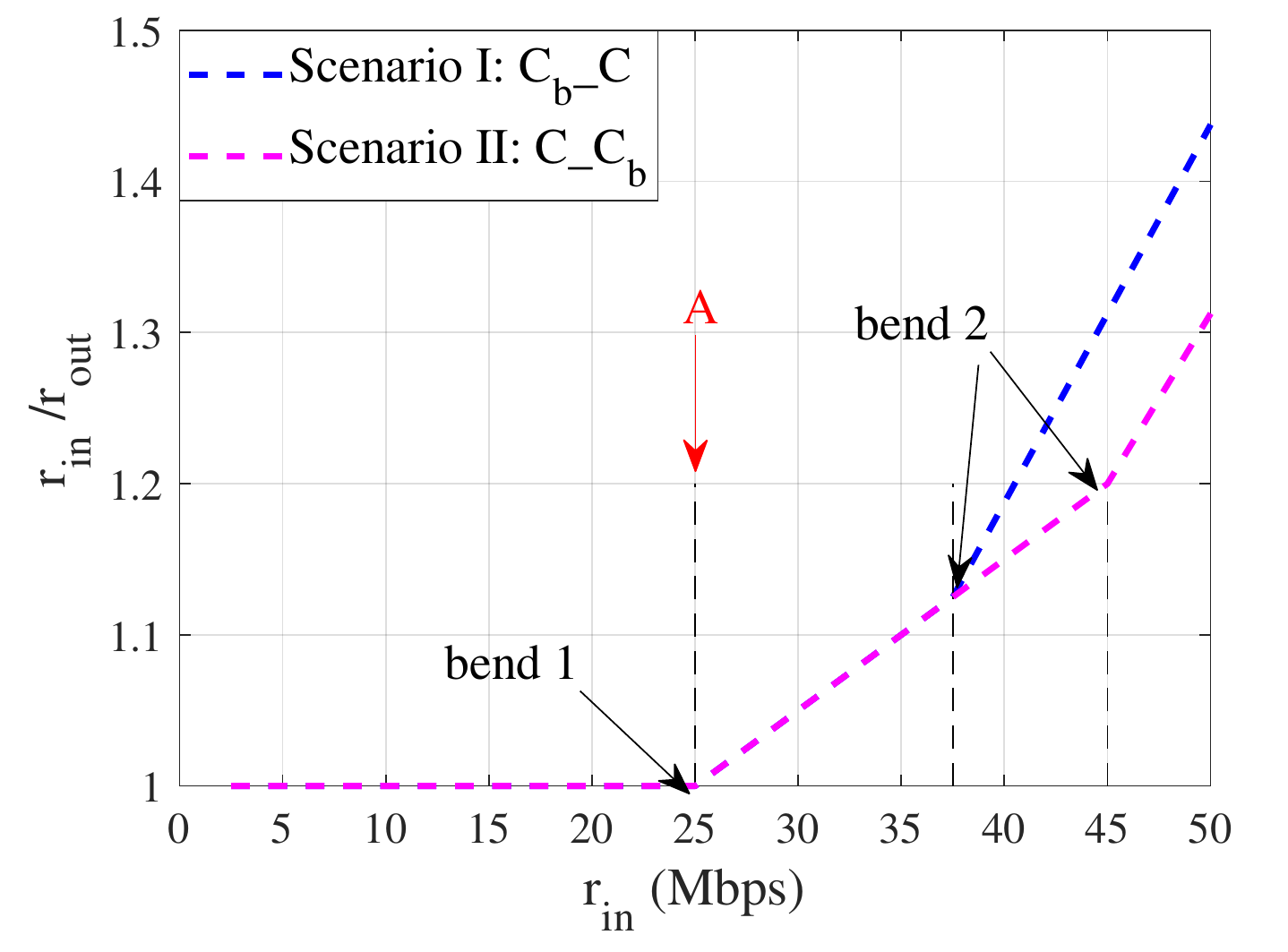}
		\label{fig:rate_res_multi}
	}
	\hfill
	\subfigure[Available bandwidth estimates for scenario I ]{
		\includegraphics[width=0.62\columnwidth]{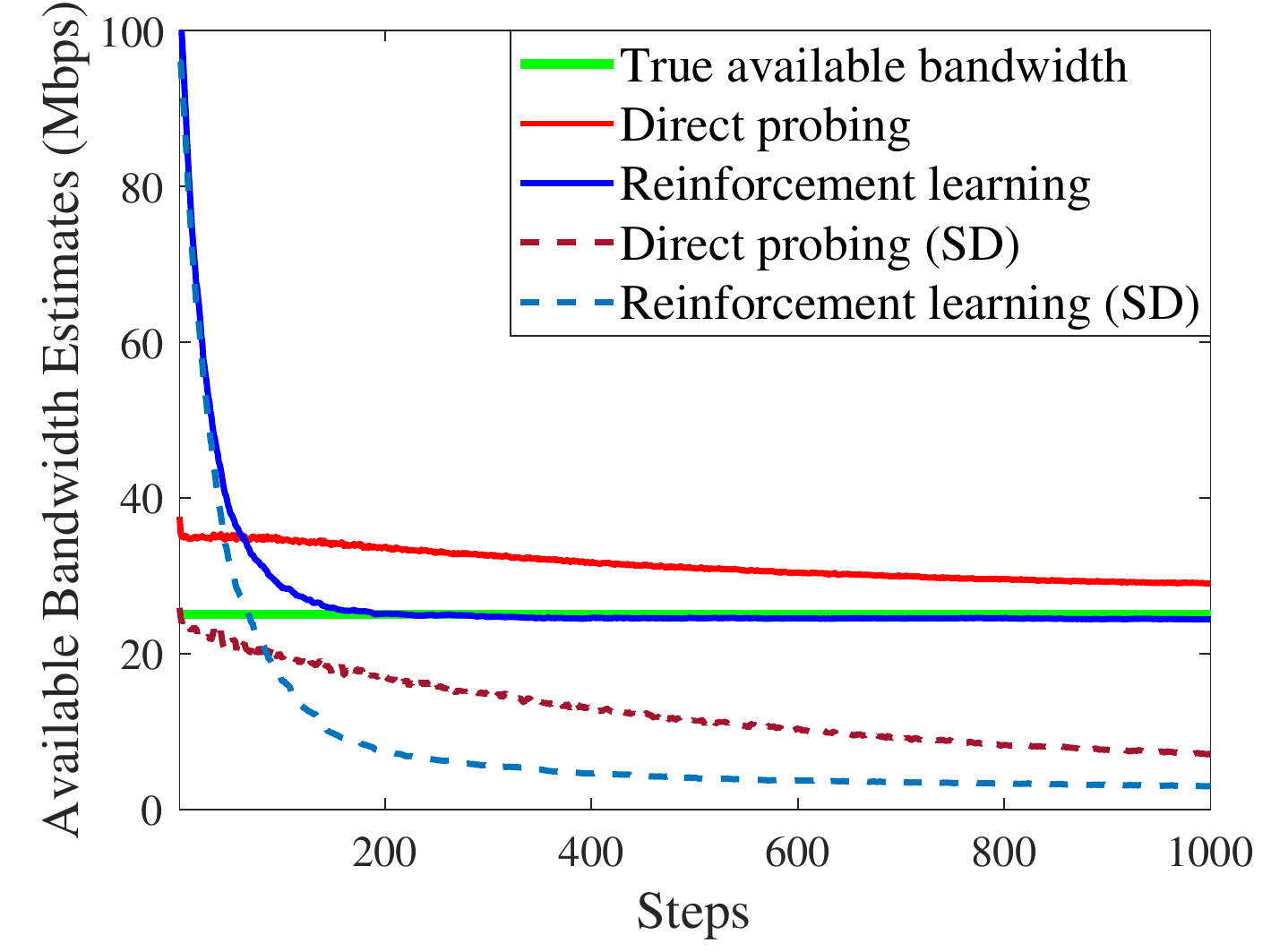}
		\label{fig:abe_b_C}
	}
	\hfill
	\subfigure[Available bandwidth estimates for scenario II]{
		\includegraphics[width=0.62\columnwidth]{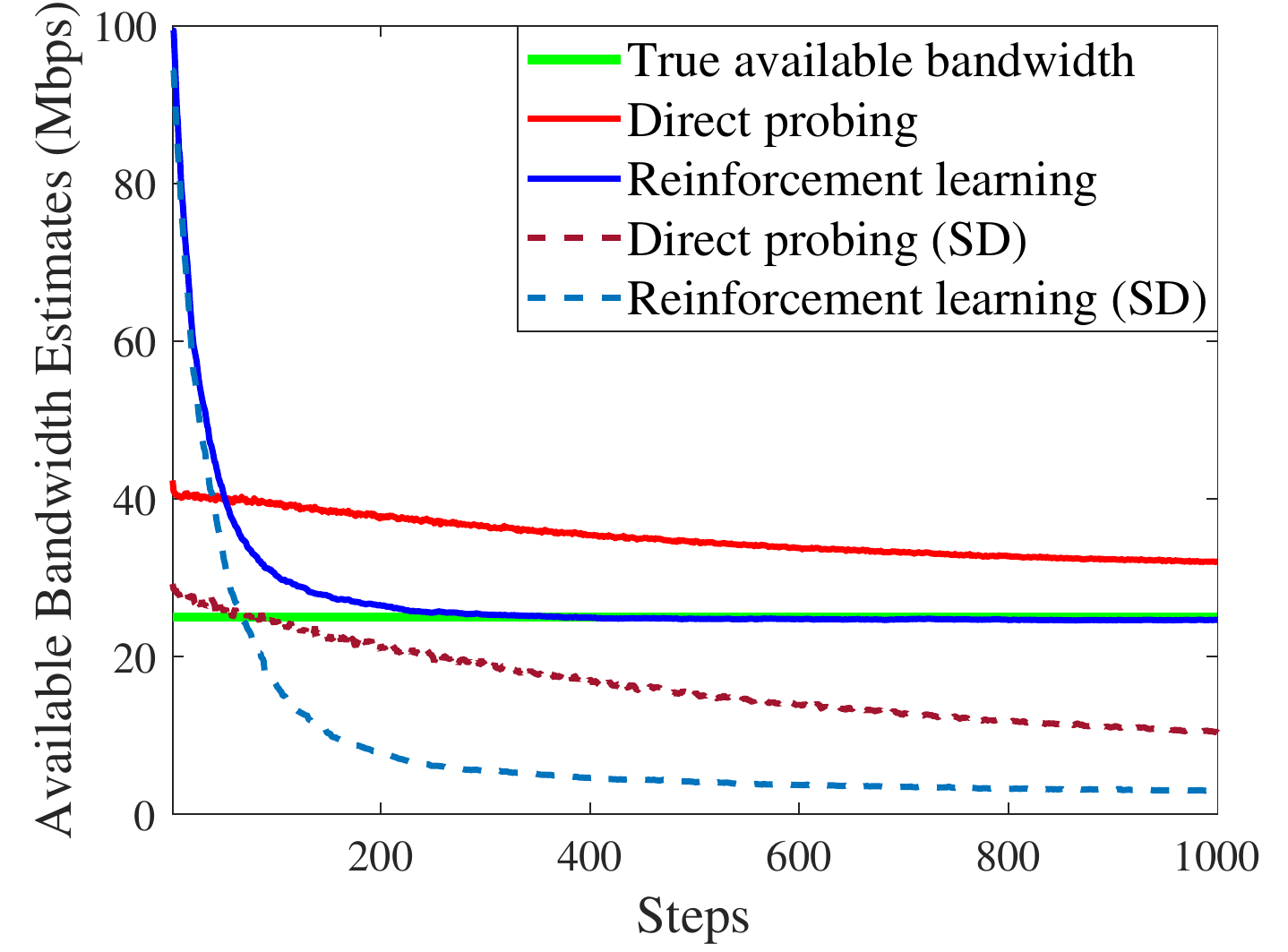}
		\label{fig:abe_C_b}
	}
	\caption{ (a) Rate response curves with two bends indicating two congestible links and average available bandwidth estimates  and their SDs  with the tight link  (b) succeeding and (c)  preceding the bottleneck link.}
\end{figure*} 

The available bandwidth estimation in multi-hop networks becomes more difficult if the tight link of a network is not the bottleneck link of the same network. As shown in Fig.~\ref{fig:tight_bottle}, link $i$ is the \emph{bottleneck link}, and link $i+1$ represents the \emph{tight link}. We investigate the available bandwidth estimation by considering two different scenarios. In Scenario I, the bottleneck link appears before the tight link. In Scenario II, the bottleneck link comes after the tight link. The existence of separate tight and bottleneck links has an impact on the shape of the rate response curve. As shown in Fig.~\ref{fig:rate_res_multi}, the curves are piece-wise linear. The two bends indicate the presence of two congestible links. We set the tight link capacity to $C=100$~Mbps and the bottleneck capacity to $C_{b}=50$~Mbps in both scenarios. We model the cross-traffic with constant bit rate $\lambda=75$~Mbps and $\lambda_{b}=12.5$~Mbps in the tight link and the bottleneck link respectively. We show the available bandwidth estimates and the corresponding SDs in Scenario I and II respectively in Fig.~\ref{fig:abe_b_C} and  Fig.~\ref{fig:abe_C_b}. The reinforcement learning-based method results in more accurate and faster estimates than the direct probing technique does. However, we observe an estimation bias in the available bandwidth estimates of the direct probing technique in both scenarios, i.e., no accurate convergence to the actual available bandwidth because of the congestion measure, $\xi$, that grows faster when the congestion occurs at both the tight and bottleneck links when the probing rates are larger than $C_{b}-\lambda_{b}=37.5$~Mbps. The effect is more noticeable in Scenario II.

%
\section{Conclusion}\label{sec:conclusion}
We have investigated how reinforcement learning can be utilized in measurement-based online available bandwidth estimation. We have proposed a method that runs $\varepsilon$-greedy algorithm to find the available bandwidth by maximizing the designated reward function. We have conducted a comprehensive measurement study in a controlled network testbed to analyze our proposed method and compare it with the piece-wise linear network model-based direct probing technique that employs a Kalman filter. Our results have shown that the reinforcement learning-based method can significantly improve the available bandwidth estimates by reducing bias and variability. The convergence of the reinforcement learning-based method is faster when compared to the direct probing technique. We have shown that our method provides better estimates in network configurations which deviate from the constant rate fluid-flow assumptions when there is cross-traffic with heavy burstiness and different intensities as well. We have further tested our method in network scenarios with multiple tight links, and in multi-hop networks where the tight link and the bottleneck link are different. We have shown that even though the additional links affect the convergence speed, the reinforcement learning-based method results in  accurate available bandwidth estimates with less variability, whereas the other method does not.
%
%
\balance
\bibliographystyle{IEEEtran}
\bibliography{IEEE}
%
%
\end{document}